\documentclass[11pt]{article}              

\usepackage[margin=25mm]{geometry}
\usepackage{graphicx}
\usepackage{subfigure}
 
\usepackage{fancyhdr}
\usepackage{amssymb,amsfonts,amsmath,amsthm}
\usepackage{natbib}
\usepackage{bstnotations}

\usepackage{authblk}
	
\graphicspath{{./},{./Figures/}}

\usepackage{bbm}
\usepackage{placeins}

  \usepackage[]{hhline}
  \usepackage{setspace}

\usepackage{lineno}
  
  \newcommand{\mathcalf}{{\mathcal F}}
  \newcommand{\card}{\text{card}}
  \newcommand{\vexi}{\ve{\xi}}
  \newcommand{\vepsi}{\ve{\psi}}
  \newcommand{\vephi}{\ve{\phi}}
  \newcommand{\matPhi}{\mat{\Phi}}
  
  \newcommand{\vevartheta}{\ve{\vartheta}}
  \newcommand{\veepsilon}{\ve{\epsilon}}
  
  \newcommand{\veXi}{\ve{\Xi}}

  \newcommand{\vey}{\ve{y}}
  \newcommand{\vez}{\ve{z}}
  \newcommand{\vea}{\ve{\alpha}}
  \newcommand{\veb}{\ve{\beta}}

  \newcommand{\matA}{\mat{A}}
  \newcommand{\matcalf}{\mathcal{F}}
  \newcommand{\matcalA}{\mathcal{A}}

\begin{document}
\title{Surrogate modelling for stochastic dynamical systems by combining NARX models and polynomial chaos expansions} 

\author[1]{C. V. Mai} \author[2]{M. D. Spiridonakos} \author[2]{E. N. Chatzi}  \author[1]{B. Sudret}

\affil[1]{Chair of Risk, Safety and Uncertainty Quantification,
  
  ETH Zurich, Stefano-Franscini-Platz 5, 8093 Zurich, Switzerland}

\affil[2]{Chair of Structural Mechanics, 
	
	ETH Zurich, Stefano-Franscini-Platz 5, 8093 Zurich, Switzerland}

\date{}
\maketitle

\abstract{The application of polynomial chaos expansions (PCEs) to the
propagation of uncertainties in stochastic dynamical models is
well-known to face challenging issues. The accuracy of PCEs degenerates
quickly in time. Thus maintaining a sufficient level of long term
accuracy requires the use of high-order polynomials.  In numerous cases,
it is even infeasible to obtain accurate metamodels with regular PCEs
due to the fact that PCEs cannot represent the dynamics.  To overcome
the problem, an original numerical approach was recently proposed that
combines PCEs and non-linear autoregressive with exogenous input (NARX)
models, which are a universal tool in the field of system
identification. The approach relies on using NARX models to mimic the
dynamical behaviour of the system and dealing with the uncertainties
using PCEs.  The PC-NARX model was built by means of heuristic genetic
algorithms.  This paper aims at introducing the least angle regression
(LAR) technique for computing PC-NARX models, which consists in solving
two linear regression problems.  The proposed approach is validated with
structural mechanics case studies, in which uncertainties arising from
both structures and excitations are taken into account. Comparison with
Monte Carlo simulation and regular PCEs is also carried out to
demonstrate the effectiveness of the proposed approach.
\\[1em] 

}

\maketitle

\section{Introduction}

Modern engineering and applied sciences have greatly benefited from the rapid increase of available computational power. Indeed, computational models allow for an accurate representation of complex physical phenomena, \eg fluid and structural dynamics.
Physical systems are in nature prone to aleatory uncertainties, such as the natural variability of system properties, excitations and boundary conditions. In order to obtain reliable predictions from numerical simulations, it is of utmost importance to take into account these uncertainties.
In this context, \emph{uncertainty quantification} has gained particular interest in the last decade. 
The propagation of uncertainties from the input to the output of the model is commonly associated with sampling-based methods, \eg Monte Carlo simulation, which are robust, however not suitable when only a small number of simulations is affordable or available. Metamodelling techniques can be employed to circumvent this issue.
Polynomial chaos expansions (PCEs) \citep{Ghanembook2003,Soize2004} are a powerful metamodelling technique that is widely used in numerous disciplines.
However, it is well-known that PCEs exhibit difficulties when being applied to stochastic dynamical models \citep{Wan2006,Gerritsma2010}.

Many efforts have been focused on solving this problem.
\citet{Wan2006,Wan2006a} proposed multi-element PCEs, which rely on the decomposition of
the random input space into sub-domains and the local computation of PCEs on each sub-element. Note that this approach is carried out in an intrusive manner, \ie the dynamical mechanism is inherently taken into account by the handling of the set of equations describing the system under investigation.
\citet{Gerritsma2010} introduced the concept of time-dependent PCEs, which consists in adding on-the-fly random variables to the set of uncertain parameters, which are the response quantities at specific time instants. 
From a similar perspective, \citet{Luchtenburg2014} proposed to combine PCEs with flow map composition, which reconstructs response time histories from short-term flow maps, each modelled with PCE.
A common feature of the approaches mentioned above is that the dynamical mechanism is mimicked by a specialized tool different from PCE.
More precisely, flow map composition and iterative enrichment in time of the random space are used to capture the dynamics of the system.

Recently, \citet{Spiridonakos2012, Spiridonakos2013, Spiridonakos2015a, Spiridonakos2015} introduced a numerical approach that is based on the 
combination of two techniques, namely PCEs and nonlinear autoregressive with exogenous input (NARX) modelling, which is a universal tool in the field of system identification.
In the proposed approach, the NARX model is used to represent the dynamical behaviour of the system, whereas PCEs tackle the uncertainty part. 
A two phase scheme is employed.
First, a \emph{stochastic} NARX model is identified to represent the dynamical system. It is characterized by a set of specified NARX model terms and associated \emph{random coefficients}. 
Second, the latter are represented as PCEs of the random input parameters which govern the uncertainties in the considered system.
In the two phases, both the NARX terms and the polynomial functions are selected by means of the heuristic genetic algorithm, which evolves randomly generated candidates toward better solutions using techniques inspired by natural evolution, \eg mutation, crossover.
The PC-NARX model is distinguished from conventional deterministic system identification tools in that it allows one to account for uncertainties arising both from
the system properties, \eg stiffness, hysteretic behaviour and energy dissipation, and from the stochastic excitations, \eg ground motions in structural analysis.
The approach proved its effectiveness in several case studies in structural dynamics \citep{Spiridonakos2015a,Spiridonakos2015}.
It is worth mentioning that early combinations of system identification tools with polynomial chaos expansions can be found in the literature.
\citet{Ghanem2005} regressed the restoring force of an oscillator on the Chebychev polynomials of state variables of the system, then used PCEs to represent the polynomial coefficients.
\citet{Wagner2007} used PC-ARIMA models with a-priori known deterministic coefficients for characterizing terrain topology. Linear ARX-PCE models were also used by \citet{Kopsaftopoulos2013}, \citet{Samara2013} and \citet{Sakellariou2016}. However, in those studies the input parameters are not characterized by known probability density functions, thus the bases are constructed from arbitrarily selected families of orthogonal polynomials. 
\citet{Jiang1993} and \citet{Poulimenos2006} used time-dependent autoregressive moving average model with stochastic parameter evolution, in which the model parameters are random variables that change with time under stochastic smoothness constraints.
Latterly, \citet{Kumar2014} represented the coefficients of a Volterra series with PCEs.
The above mentioned approaches were carried out in the time domain. In the frequency domain, \citet{Pichler2009} used linear regression-based metamodels to represent the frequency response functions of linear structures. In particular, the same strategy was recently used with PCEs by \citet{Yang2015} and \citet{Jacquelin2015a}.

In summary, the PC-NARX model consists of two components, namely a NARX and a PCE model. \citet{Spiridonakos2015,Spiridonakos2015a} computed the two components with the heuristic genetic algorithm. This paper aims at introducing the so-called least angle regression (LARS) technique \citep{Efron2004} for the computation of both NARX and PCE which are merely linear regression models. Indeed LARS has proven to be efficient in computing adaptive sparse PCEs at a relatively low computational cost \citep{Blatman2011b}. LARS has been recently used for selecting the NARX terms in the context of system identification \citep{Zhang2015b}.
Yet the original contribution of this paper is to use LARS as a selection scheme for both NARX terms and PCE basis terms. This way we provide a new fully non-intrusive surrogate modelling technique that allows to tackle nonlinear dynamical systems with uncertain parameters.

The paper is organized as follows: in the first two sections, the theory of PCEs and NARX model are briefly recalled. Section 4 presents the PC-NARX model and the proposed LARS-based approach for computing it.
The methodology is illustrated with three benchmark engineering case studies in Section \ref{sec5}. Detailed discussions on the approach are then given before the final conclusions.

\section{Polynomial chaos expansions}


\subsection{Polynomial chaos expansions}
\label{sec2.1}
Let us consider the computational model $Y=\cm(\veXi)$ where $\veXi=(\Xi_1 \enum \Xi_M)$ is a $M$-dimensional input vector of random variables with given joint probability density function $f_{\veXi}$ defined over an underlying probability space $(\varOmega, \matcalf, \mathbb{P})$ and $\cm:\, \vexi  \in \cd_{\veXi} \subset \Rr^M \mapsto \Rr$ is the computational model of interest, where $\cd_{\veXi}$ is the support of the distribution of $\veXi$. Without loss of generality, we assume that the input random variables are independent.
When dependent input variables are to be considered, an iso-probabilistic transform
is used to map the latter to independent auxiliary variables \citep{Sudret2015}.

The scalar output $Y$, which is assumed to be a second order random variable, \ie $\Esp{Y^2} < +\infty$, can be represented using generalized polynomial chaos expansion as follows \citep{Xiu2002, Soize2004}:
\begin{equation}
 Y= \sum\limits_{\vea \in \Nn^{M}} y_{\vea} \ve{\psi}_{\vea}(\veXi), 
 \label{eq2.1.1}
\end{equation}
in which $\ve{\psi}_{\vea}(\veXi) = \prod\limits_{i=1}^{M} \psi_{\alpha_i}^i(\Xi_i)$ are multivariate \emph{orthonormal} polynomials obtained by the tensor product of univariate polynomials $\psi_{\alpha_i}^i(\Xi_i)$, $y_{\vea}$ are associated deterministic coefficients, $\vea = \prt{\alpha_1 \enum \alpha_M}$ is the multi-index vector with $\alpha_i, i=1\enum M$ being the the degree of the univariate polynomial $\psi_{\alpha_i}^i(\Xi_i)$. The latter constitutes a basis of orthonormal polynomials with respect to the marginal probability measure.
For instance, when $\Xi_i$ is a uniform (resp. standard normal) random variable, the corresponding basis comprises orthonormal Legendre (resp. Hermite) polynomials \citep{Abramowitz}. 

In practice, it is not tractable to use an infinite series expansion. An approximate representation is obtained by means of a truncation:
\begin{equation}
 Y = \sum\limits_{\vea \in \matcalA} y_{\vea} \ve{\psi}_{\vea}(\veXi) + \epsilon \equiv \cm^{PC}(\veXi) + \epsilon,
 \label{eq2.2.2}
\end{equation}
in which $\matcalA$ is the truncation set and $\epsilon$ is the truncation-induced error. \citet{Blatman2011b} introduced a hyperbolic truncation scheme, which consists in selecting all polynomials satisfying the following criterion:
\begin{equation}
 \matcalA_q^{M,p} = \acc{ \vea \in \Nn^M: \quad \norme{\vea}{q} \eqdef  \prt{\sum\limits_{i=1}^{M} \alpha_i^q }^{1/q}  \leq p },
\end{equation}
with $p$ being the highest total polynomial degree, $0 < q \leq 1$ being the parameter determining the hyperbolic truncation surface.
To further reduce the number of candidate polynomials, one can additionally apply a low-rank truncation scheme which reads \citep{BlatmanThesis}: 
\begin{equation}
	 \matcalA_{q,r}^{M,p} =\acc{ \vea \in \Nn^M: \quad \norme{\vea}{0} \eqdef \sum\limits_{i=1}^M \mathbbm{1}_{\alpha_i > 0} \leq r, \, \norme{\vea}{q} \leq p },
\end{equation}
where $\norme{\vea}{0}$ is the rank of the multivariate polynomial ${\psi}_{\vea}$, defined as the total number of non-zero components $\alpha_i, i= 1 \enum M$. The prescribed rank $r$ is usually chosen as a small integer value, \eg $r= 2,3.$

\subsection{Computing the coefficients by least square minimization}
The computation of the coefficients in Eq.~\eqref{eq2.2.2} can be conducted by means of intrusive approach (\ie Galerkin scheme) or non-intrusive approaches (\eg stochastic collocation, projection, regression methods) \citep{SudretHDR,XiuBook2010}.
The regression method consists in estimating the set of coefficients $\hat{\vey}_{\vea} = \acc{y_{\vea}, \vea \in \matcalA}$ that minimizes the mean square error:
\begin{equation}
	\Esp{\epsilon^2} \eqdef \Esp{ \prt{Y - \sum\limits_{\vea \in \matcalA} y_{\vea} {\vepsi}_{\vea}(\veXi)}^2 },
\end{equation}
which means:
\begin{equation}
	\hat{\vey}_{\vea} = \arg \min \limits_{ \vey_{\vea} \in \Rr^{\card \matcalA}} 
	\Esp{ \prt{\cm(\veXi) - \sum\limits_{\vea \in \matcalA} y_{\vea} {\vepsi}_{\vea}(\veXi)}^2 }.
\end{equation}
In practice, the coefficients are obtained by minimizing an empirical mean over a sample set:
\begin{equation}
	\hat{\vey}_{\vea} = \arg \min \limits_{ \vey_{\vea} \in \Rr^{\card \matcalA}} 
		\dfrac{1}{N} \sum\limits_{i=1}^{N}{ \prt{\cm(\vexi^{(i)}) - \sum\limits_{\vea \in \matcalA} y_{\vea} {\vepsi}_{\vea}(\vexi^{(i)})}^2 },
		\label{eq2.2.3}
\end{equation}
%
where $\cx = \acc{\vexi^{(i)}, \, i = 1 \enum N}$ is an experimental design (ED) obtained with random sampling 
of the input random vector. The computational model $\cm$ is run for each point of the ED, yielding the vector of output values $\cy = \acc{y^{(i)} = \cm(\vexi^{(i)}), \, i = 1 \enum N}$.
By evaluating the polynomial basis onto each sample point in the ED, one obtains the information matrix, which is
defined as follows:
\begin{equation}
	\matA \eqdef \nobreak \acc{A_{ij} = \vepsi_j(\vexi^{(i)}), \, i = 1 \enum N, \, j = 1\enum \card  \matcalA} ,
\end{equation}
\ie the $i^{\text{th}}$ row of $\matA$ is the evaluation of the polynomial basis functions at the point $\vexi^{(i)}$ in the ED.
Eq.~\eqref{eq2.2.3} basically represents the problem of estimating the parameters of a linear regression model, for which the least squares solution reads:
\begin{equation}
	\hat{\vey}_{\vea}  = \prt{ \matA\tr \, \matA }^{-1} \, \matA\tr \, \cy.
\end{equation}

\subsection{Error estimator}
\label{sec2.2}

Herein, the accuracy of the truncated expansion (Eq.~\eqref{eq2.2.2}) is estimated by means of the leave-one-out (LOO) cross validation technique, which allows a fair error estimation at an affordable computational cost \citep{Blatman2010b}. 
Assume that one is given an experimental design (ED) $\cx = \acc{\vexi^{(i)}, \, i = 1 \enum N}$ and the associated vector of model responses $\cy = \acc{y^{(i)}, \, i = 1 \enum N}$ obtained by running the computational model $\cm$.
Cross-validation consists in partitioning the ED into two complementary subsets, training a model using one subset, then validating its prediction on the other subset. In this context, the term LOO means that the validation set comprises only one sample and the model is computed using all the $N-1$ remaining samples. 
A single point of the experimental design is left out at a time, a PCE $\cm^{\text{PC}\backslash i}$ is built using the remaining points and the corresponding prediction error on the validation point is computed:
\begin{equation}
	\Delta_i \eqdef \cm(\vexi^{(i)}) - \cm^{\text{PC}\backslash i}(\vexi^{(i)}).
\end{equation}
 The LOO error is defined by:
\begin{equation}
 	\widehat{\text{Err}}_{LOO} = \dfrac{1}{N} \sum\limits_{i=1}^{N} \Delta_i^2 = 
 	\dfrac{1}{N} \sum\limits_{i=1}^{N} \prt{ \cm(\vexi^{(i)}) - \cm^{\text{PC}\backslash i}(\vexi^{(i)})}^2.
\end{equation}

At first glance, computing the LOO error may appear computationally demanding since it requires to train and validate $N$ different PCE models. However, by means of algebraic derivations, one can compute the LOO error $\widehat{\text{Err}}_{LOO}$ from a \emph{single} PCE model $\cm^{\text{PC}}(\cdot)$ built with the full ED $\cx$ as follows \citep{BlatmanThesis}:
\begin{equation}
	\widehat{\text{Err}}_{LOO} =  \dfrac{1}{N} \sum\limits_{i=1}^{N} \prt{ \dfrac{ \cm(\vexi^{(i)}) - \cm^{\text{PC}}(\vexi^{(i)}) }{ 1 - h_i} }^2,
	\label{eq12}
\end{equation}
in which $h_i$ is the $i^{\text{th}}$ diagonal term of the matrix $\matA \, \prt{ \matA\tr \matA }^{-1} \matA\tr $.

\subsection{Adaptive sparse PCEs based on least angle regression}
\label{sec2.3}
To obtain satisfactory results with the least-square minimization method, the number of model evaluations $N$ is commonly required to be two to three times the cardinality $P = \card \matcalA$ of the polynomial chaos basis \citep{BlatmanThesis}. In case when the problem involves high dimensionality and when high order polynomials must be used, \ie $P$ is large, the required size of the ED soon becomes excessive. An effective scheme that allows one to obtain accurate polynomial expansions with limited number of model evaluations is of utmost importance.

In the following, we will shortly describe the adaptive sparse PCE scheme proposed by \citet{Blatman2011b} which is a non-intrusive least-square minimization technique based on the least angle regression (LARS) algorithm \citep{Efron2004}. 
LARS is an efficient numerical technique for variable selection in high dimensional problems. It aims at detecting 
the most relevant predictors among a large set of candidates given a limited number of observations. In the context of PCEs, LARS allows one to achieve a sparse representation, \ie the number of retained polynomial functions is small compared to the size of the candidate set. This is done in an adaptive manner, \ie the candidate functions become active one after another in the descending order of their importance. The relevance of a candidate polynomial is measured by means of its correlation with the current residual of the expansion obtained in the previous iteration. The optimal PCE is chosen so as to minimize the leave-one-out error estimator in Eq.~\eqref{eq12}.
In particular, LARS generally requires a relatively small ED, thus being effective in realistic applications.
The reader is referred to \citet{Efron2004} for more details on the LARS technique and to \citet{Blatman2011b} for its implementation in the adaptive sparse PCE scheme.

\subsection{Time-frozen PCEs and associated statistics}

In this paper, we focus on stochastic dynamical systems where the random output response is a time dependent quantity $Y(t) = \cm(\veXi,t)$. In this context, the polynomial chaos representation of the response may be cast as:
\begin{equation}
 Y(t) = \sum\limits_{\vea \in \matcalA} y_{\vea}(t) \ve{\psi}_{\vea}(\veXi) + \epsilon(t),
 \label{eq2.4.1}
\end{equation}
in which the notation $y_{\vea}(t)$ indicates time-dependent PCE coefficients and $\epsilon(t)$ is the residual at time $t$. The representation of a time-dependent quantity by means of PCEs as in Eq.~\eqref{eq2.4.1} is widely used in the literature, see \eg \citet{Pettit2006,LeMaitre2009,Gerritsma2010}. 
At a given time instant $t$, the coefficients $\acc{y_{\vea}(t),\vea \in \matcalA}$ and the accuracy of the PCEs may be estimated by means of the above mentioned techniques (see Sections~\ref{sec2.2} and \ref{sec2.3}).
In such an approach, the metamodel of the response is computed \emph{independently} at each time instant, hence the name time-frozen PCEs. This naive approach is presented here for the sake of comparison with the PC-NARX method introduced later.

Eq.~\eqref{eq2.4.1} can be used to compute the evolution of the response statistics. The multivariate polynomial chaos functions are orthonormal, \ie:
\begin{equation}
	\Esp{ \ve{\psi}_{\vea}(\veXi) \, \ve{\psi}_{\veb}(\veXi) }  \eqdef \int\limits_{\cd_{\veXi}} \ve{\psi}_{\vea}(\vexi) \, \ve{\psi}_{\veb}(\vexi) \, f_{\veXi}(\vexi) \, \di \vexi = \delta_{\vea \veb} \quad \forall \vea, \, \veb \in \Nn^M,
\end{equation}
in which $\delta_{\vea \veb}$ is the Kronecker symbol that is equal to 1 if $\vea = \veb$ and equal to 0 otherwise. In particular, each multivariate polynomial is orthonormal to $\ve{\psi}_{\ve{0}}(\veXi) = 1$, which means
$\Esp{\ve{\psi}_{\vea}(\veXi)}  = 0 \; \forall \vea \neq \ve{0}$ and $\Var{\ve{\psi}_{\vea}(\veXi)} = \Esp{ \prt{\ve{\psi}_{\vea}^2(\veXi)} } = 1 \; \forall \vea \neq \ve{0}$.
Thus, the time dependent mean and standard deviation of the response can be estimated with no additional cost  by means of the post-processing of the truncated PC coefficients in Eq.~\eqref{eq2.4.1} as follows:
\begin{equation}
	\hat{\mu}_{Y(t)} \eqdef \Esp{ \sum\limits_{\vea \in \matcalA} y_{\vea}(t) \ve{\psi}_{\vea}(\veXi)} = y_{\ve{0}}(t) \; ,
\end{equation}

\begin{equation}
	\hat{\sigma}_{Y(t)}^2 \eqdef \Var{ \sum\limits_{\vea \in \matcalA} y_{\vea}(t) \ve{\psi}_{\vea}(\veXi) } = \sum\limits_{\substack{{\vea \in \matcalA} \\ \vea \neq \ve{0} }} y_{\vea}^2(t).
\end{equation}

\section{Nonlinear autoregressive with exogenous input model}

Let us consider a computational model $y(t) = \cm(x(t))$ where $x(t)$ is the time-dependent input excitation and $y(t)$ is the response time history of interest. 
System identification aims at building a mathematical model describing $\cm$ using the observed data of the input and output signals. In this paper, we focus on system identification in the time domain. 
One discretizes the time duration under investigation in $T$ discrete instants $t = 1 \enum T$.
A nonlinear autoregressive with exogenous input (NARX) model allows one to represent the output quantity at a considered time instant as a function of its past values and values of the input excitation at the current or previous instants \citep{Chen1989, Billings2013}:
	\begin{equation}
		y(t) =  \mathcalf \prt{\vez(t)} + \epsilon(t) = \mathcalf \prt{x(t) \enum x(t-n_x), y(t-1) \enum y(t-n_y)}  + \epsilon_t ,
	\end{equation}
where $\mathcalf(\cdot)$ is the underlying mathematical model to be identified, $\ve{z}(t) = (x(t) \enum x(t-n_x), y(t-1) \enum y(t-n_y))\tr$ is the vector of current and past values, $n_x$ and $n_y$ denote the maximum input and output time lags,
$\epsilon_t \sim \cn(0,\sigma_{\epsilon}^2(t))$ is the residual error of the NARX model. In standard NARX models, the residuals are assumed to be independent normal variables with zero mean and variance $\sigma_{\epsilon}^2(t)$.
There are multiple options for the mapping function $\mathcalf(\cdot)$. In the literature, the following linear-in-the-parameters form is commonly used:
		\begin{equation}
			y(t) = \sum\limits_{i=1}^{n_g} \vartheta_i \, g_i( \ve{z}(t) ) + \epsilon_t ,
			\label{eq3.1}
		\end{equation}
in which $n_g$ is the number of model terms $g_i( \ve{z}(t) )$ that are functions of the regression vector $\ve{z}(t)$ and  $\vartheta_i$ are the coefficients of the NARX model.
  
Indeed, a NARX model allows ones to capture the dynamical behaviour of the system which follows the principle of causality, \ie the current output quantity (or state of the system) $y(t)$ is affected by its previous states $\acc{y(t-1) \enum y(t-n_y)}$ and the external excitation $\acc{x(t) \enum x(t-n_x)}$. Note that the cause-consequence effect tends to fade away as time evolves, therefore it suffices to consider only a limited number of time lags before the current time instant. It is worth emphasizing that the model terms may be constructed from a variety of global or local basis functions. For instance the use of polynomial NARX model with $g_i( \ve{z}(t) )$ being polynomial functions is relatively popular in the literature.

The identification of a NARX model for a system consists of two major steps. The first one is \emph{structure selection}, \ie determining
which NARX terms $g_i( \ve{z}(t) )$ are in the model. The second step is parameter estimation, \ie determining the associated model coefficients. 
Note that structure selection, particularly for systems involving nonlinearities, is critically important and difficult. Including spurious terms in the model
leads to numerical and computational problems \citep{Billings2013}.
Billings suggests to identify the simplest model to represent the underlying dynamics of the system, which can be achieved by using the orthogonal least squares algorithm and its derivatives to select the relevant model terms one at a time \citep{Billings2013}.
Different approaches for structure selection include trial and error methods, see \eg \citet{Chen2011,Piroddi2008}, and correlation-based methods, see \eg \citet{Billings2008, Wei2008a}.
There is a rich literature dedicated to this topic, however discussions on those works are not in the scope of the current paper.

The identified model can be used for several purposes. First, it helps the analysts reveal the mechanism and behaviour of the underlying system. Understanding how a system operates offers one the possibility to control it better. Second, the identified mathematical model can be utilized for predicting future responses of the system. From this point of view, it can be considered a metamodel (or approximate model) of the original $\cm$.

\section{Polynomial chaos - nonlinear autoregressive with exogenous input model}

Consider a computational model $y(t,\vexi) = \cm(x(t,\vexi_x), \vexi_s)$ where
$\vexi = \prt{\vexi_x, \, \vexi_s}\tr$ is the vector of uncertain parameters, $\vexi_x$ and $\vexi_s$ respectively represent the uncertainties in the input excitation $x(t,\vexi_x)$ and in the system itself. For instance, $\vexi_x$ can contain parameters governing the amplitude and frequency content of the excitation time series, while $\vexi_s$ can comprise parameters determining the system properties such as geometries, stiffness, damping and hysteretic behaviour.

\citet{Spiridonakos2015a, Spiridonakos2015} proposed a numerical approach based on PCEs and NARX model to identify the metamodel of such a dynamical system with uncertainties arising from both the excitation and the system properties.
The time-dependent output quantity is first represented by means of a NARX model:
	\begin{equation}
		y(t,\vexi) =  \sum\limits_{i=1}^{n_g} \vartheta_i(\vexi) \, g_i( \ve{z}(t) ) + \epsilon_{g}(t,\vexi) ,
		\label{eq4.1}
	\end{equation}
in which the model terms $g_i\prt{\ve{z}(t)}$ are functions of the regression vector
$\ve{z}(t) = (x(t) \enum x(t-n_x),$ $ y(t-1) \enum y(t-n_y) )\tr$, $n_x$ and $n_y$ denote the maximum input and output time lags, $\vartheta_i(\vexi)$ are the coefficients of the NARX model, $\epsilon_{g}(t,\vexi)$ is the residual error, with zero mean Gaussian distribution and variance $\sigma_{\epsilon}^2(t)$.
The proposed NARX model differs from the classical NARX model in the fact that the coefficients $\vartheta_i(\vexi)$ are functions of the uncertain input parameters $\vexi$ instead of being deterministic.
The stochastic coefficients $\vartheta_i(\vexi)$ of the NARX model are then represented by means of truncated PCEs as follows \citep{Soize2004}:
\begin{equation}
	\vartheta_i(\vexi) = \sum\limits_{j=1}^{n_{\psi}} \vartheta_{i,j} \, \psi_j(\vexi) + \epsilon_i,
	\label{eq4.2}
\end{equation}
in which $\acc{\psi_j(\vexi), j = 1 \enum n_{\psi}}$ are multivariate orthonormal polynomials of $\vexi$, $\left\{ \vartheta_{i,j}, i = 1 \enum n_g, \right.$ $ä\left. j = 1 \enum n_{\psi}\right\}$ are associated PC coefficients and $\epsilon_i$ is the truncation error.
Finally, the PC-NARX model reads:
\begin{equation}
	y(t,\vexi) = \sum\limits_{i=1}^{n_g} \sum\limits_{j=1}^{n_{\psi}} \vartheta_{i,j} \, \psi_j(\vexi) \, g_i( \ve{z}(t) ) + \epsilon(t,\vexi),
	\label{eq4.3}
\end{equation}
where $\epsilon(t,\vexi)$ is the total error time series due to the truncations of NARX and PCE models. 
In the proposed approach, the NARX model is used to capture the dynamics of the considered system, whereas PCEs are used to propagate uncertainties. 

Let us discuss the difference between the PC-NARX model and the conventional time-dependent PCE formulation in Eq.~\eqref{eq2.4.1}. For the sake of clarity, Eq.~\eqref{eq4.3} can be rewritten as follows:
\begin{equation}
	y(t,\vexi) = \sum\limits_{j=1}^{n_{\psi}} \prt{\sum\limits_{i=1}^{n_g}  \vartheta_{i,j} \, g_i( \ve{z}(t) ) } \psi_j(\vexi) \,  + \epsilon(t,\vexi).
	\label{eq4.4}
\end{equation}
At a considered instant $t$, the polynomial coefficients $y_j(t) \eqdef \sum\limits_{i=1}^{n_g}  \vartheta_{i,j} \, g_i( \ve{z}(t) )$ are represented as functions of the past values of the excitation and the output quantity of interest. Consequently the polynomial coefficients follow certain dynamical behaviours. In the conventional model in Eq.~\eqref{eq2.4.1}, the polynomial coefficients at time $t$ are \emph{deterministic}. Therefore \emph{high and increasing} polynomial order is required to maintain an accuracy level and properly capture the dynamics as time evolves \citep{Wan2006}. In contrast, when a functional form is used to relate the coefficients $y_j(t)$ with the excitation and output time series, \emph{constant and low} polynomial order suffices. \citet{Spiridonakos2015} used PC-NARX models with fourth order PCEs to obtain remarkable results in the considered structural dynamics case studies. In the literature, \citet{Gerritsma2010} showed that when applying time-dependent PCEs, \ie adding previous responses to the set of random variables to represent current response, low-order polynomials could also be used effectively. From a similar perspective, the PC-flow map composition scheme proposed by \citet{Luchtenburg2014} was also proven efficient in solving the problems with low polynomial order, which was impossible with PCEs alone.

Indeed, not all the NARX and PC terms originally specified are relevant, as commonly observed in practice. The use of redundant NARX or PC terms might lead to large inaccuracy. Therefore, it is of utmost importance to identify the correct structure of NARX and PC models, \ie to select appropriate NARX terms and PC bases. To this end, \citet{Spiridonakos2015} proposed a two-phase approach, in which the NARX terms and PC functions are subsequently selected by means of the genetic algorithm. However, due to the linear-in-parameters formulations of the NARX model (Eq.~\eqref{eq4.1}) and the PC expansions (Eq.~\eqref{eq4.2}),
the question of selecting NARX and PC terms boils down to solving two linear regression problems. To this end, it appears that one can use techniques that are specially designed for linear regression analysis, for instance least angle regression (LARS) \citep{Efron2004}.
LARS has been recently used for selecting NARX terms \citep{Zhang2015b}. 
The use of LARS in the field of system identification can be classified as a correlation-based method, which selects the NARX terms that make significant contribution to the output using
correlation analysis, see \eg \citet{Billings2008, Wei2008a}.
LARS has also been used in the adaptive sparse PCE scheme and proven great advantages compared to the other predictor selection methods, \ie fast convergence and high accuracy with an ED of limited size.

\subsection{Least angle regression-based approach}
In this section, we introduce least angle regression (LARS) for the selection of appropriate NARX and PCE models. A two phase approach is used, which sequentially selects NARX and PCE models as follows:
\begin{itemize}
	\item Phase 1: Selection of the appropriate NARX model among a set of candidates.
		\begin{itemize}
			\item Step 1.1: One specifies general options for the NARX model (model class and related properties), \eg type of basis functions (polynomials, neural network, sigmoid functions, \etc), maximum time lags of input and output, properties of the basis functions (\eg maximum polynomial order). Note that it is always preferable to start with simple models having a reasonable number of terms. In addition, any available knowledge on the system, \eg number of degrees of freedom, type of non-linear behaviour, should be used in order to obtain useful options for the general NARX structure. This leads to a full NARX model which usually contains more terms than actually needed for a proper representation of the considered dynamical system.  At this stage, one assumes that the specified full NARX model contains the terms that can sufficiently describe the system. This assumption will be verified in the final step of this phase. 
		     \item Step 1.2: One selects some candidate NARX models being subsets of the specified full model. To this end, one considers the experiments exhibiting a high level of non-linearity. For instance those experiments can be chosen with measures of nonlinearity or by inspection of the simulations with maximum response values exceeding a specified threshold.
		     For each of the selected experiments, one determines a candidate NARX model containing a subset of the NARX terms specified by the full model. This is done using LARS and the input-output time histories of the considered experiment. 
		     
		     For experiment \#$k$, the one-step-ahead prediction of the response reads:
		     \begin{equation}
		     	\hat{y}_p(t,\vexi_k) = \sum\limits_{i=1}^{n_g}  \vartheta_{i}(\vexi_k)  \, g_i( \hat{\ve{z}}_p(t) ) ,
		     	\label{eq1sap}
		     \end{equation}
		     in which
		     \begin{equation}
		     	\hat{\ve{z}}_p(t) = \prt{x(t) \enum x(t-n_x), y(t-1) \enum y(t-n_y)}\tr.
		     \end{equation}
		     It is worth emphasizing that $y(t)$ is the recorded data used for training the NARX model.
		      Denoting $\vephi(t) = \acc{ g_i( \hat{\ve{z}}_p(t), \, i = 1 \enum n_g )}\tr$ and $\vevartheta = \acc{\vartheta_i, \, i = 1 \enum n_g}\tr$,
		     the residual time series reads:
		     \begin{equation}
		     	\epsilon_p(t,\vexi_k)  = y(t,\vexi_k) - \hat{y}_p(t,\vexi_k) = y(t,\vexi_k) - \vephi\tr(t) \, \vevartheta(\vexi_k).
		     	\label{eqPE}
		     \end{equation}
		     Thus, the sum of squared errors is given by:
		     		     			 \begin{equation}
		     		     				   \sum\limits_{t=1}^{T} \bra{ \epsilon_p(t,\vexi_k) }^2   =  \sum\limits_{t=1}^{T} \bra{ y(t,\vexi_k) - \vephi\tr(t) \, \vevartheta(\vexi_k) }^2.
		     		     				   \label{eqssqrPE}
		     		     			 \end{equation}
		     Using Eq.~\eqref{eqPE} and assembling all time instants in the $k$-th experiment, one obtains:
		     \begin{equation}
		     	\left[ \begin{array}{c}
		     					y(1,\vexi_k)   \\
		     				\vdots \\
		     					y(T,\vexi_k)
		     				\end{array} \right]
		     				=
		     	\left[ \begin{array}{c}
		     						\vephi\tr(1)   \\
		     					\vdots \\
		     						\vephi\tr(T)
		     					\end{array} \right]	\, \vevartheta(\vexi_k) + 
		     	\left[ \begin{array}{c}
		     							\epsilon_p(1,\vexi_k)   \\
		     						\vdots \\
		     							\epsilon_p(T,\vexi_k)
		     						\end{array} \right]						
		     \end{equation}
		     The above equation can be rewritten in matrix notations as follows:
		     \begin{equation}
		     	\vey_k = \matPhi_k \, \vevartheta(\vexi_k) + \veepsilon_k ,
		     	\label{eqOLS}
		     \end{equation}
		     where $\vey_k$ is the $T \times 1$ vector of output time-series, $\matPhi_k$ is the $T \times n_g$ information matrix with the $i^{th}$ row containing the evaluations of NARX terms $\vephi(t)$ at instant $t=i$ and $\veepsilon_k$ is the residual vector.
		     This is typically the equation of a linear regression problem, for which the relevant NARX regressors among the NARX candidate terms $\vephi(t)$ can be selected by LARS.

		     Note that the same candidate model might be obtained from different selected experiments.
		     In theory all the available experiments can be considered, \ie the number of candidate models is at most the size of the ED. Herein, we search for the appropriate NARX model among a limited number of experiments which are exhibiting strong non-linearity.

			 \item Step 1.3: For each candidate NARX model, the corresponding NARX coefficients are computed for each of the experiments by ordinary least-squares. 
			 The parameters $\vevartheta(\vexi_k)$ minimizing the total errors in Eq.~\eqref{eqssqrPE} is the least-squares solution
			 		     of Eq.~\eqref{eqOLS}, \ie:
			 		     \begin{equation}
			 		     	\vevartheta(\vexi_k) = \arg \underset{\ve{\vartheta}}{\min} (\veepsilon_k\tr \, \veepsilon_k) = \bra{ \matPhi_k\tr \, \matPhi_k }^{-1} \, \matPhi_k\tr \, \vey_k,
			 		     	\label{eqOLSsolution}
			 		     \end{equation}
			 with the information matrix $\matPhi_k$ containing only the NARX regressors specified in the NARX candidate.
			 Having at hand the NARX coefficients, the free-run reconstruction for each experiment output is conducted as follows:
			 \begin{equation}
			 		     	\hat{y}_s(t,\vexi_k) = \sum\limits_{i=1}^{n_g}  \vartheta_{i}(\vexi_k)  \, g_i( \hat{\ve{z}}_s(t) ) ,
			 		     \end{equation}
			 		     in which
			 		     \begin{equation}
			 		     	\hat{\ve{z}}_s(t) = \prt{x(t) \enum x(t-n_x), \hat{y}_s(t-1) \enum \hat{y}_s(t-n_y)}\tr.
			 		     \end{equation}
			 It is worth underlining that the free-run reconstruction of the response is obtained using only the excitation time series $x(t)$ and the response initial condition $y_0$. The response is reconstructed recursively, \ie its estimate at one instant is used to predict the response at later instants. This differs from Eq.~\eqref{eq1sap} where the recorded response was used in the recursive formulation.
			 The relative error for simulation \#$k$ reads:
			 			 	 \begin{equation}
			 			 	 	\epsilon_k = \dfrac{ \sum\limits_{t=1}^{T} (y(t, \vexi_k) - \hat{y}_s(t, \vexi_k))^2 }{\sum\limits_{t=1}^{T} (y(t, \vexi_k) - \bar{y}(t, \vexi_k) )^2},
			 			 	 	\label{eqSE}
			 			 	 \end{equation}
			 in which $\hat{y}_s(t, \vexi_k)$ is the output trajectory reconstructed by the NARX model and $\bar{y}(t, \vexi_k)$ is the mean value of the response time series $y(t,\vexi_k)$.
			 \item Step 1.4: One selects the most appropriate NARX model among the candidates. 
			 Herein, the error criterion of interest is the mean value of the relative errors:
			 			 \begin{equation}
			 				 \bar{\epsilon}  = \dfrac{1}{K} \sum\limits_{k=1}^{K} \epsilon_k .
			 				 \label{eq4.3.3}
			 			 \end{equation}
			 We propose to choose the model that achieves a sufficiently small overall error on the conducted experiments, \eg $\bar{\epsilon} < 1 \times 10^{-3}$, with the smallest number of NARX terms. In other words, the appropriate model is the simplest one that allows to capture the system dynamical behaviour, thus following the principle suggested by Billings \cite{Billings2013}.

To refine the estimated coefficients, a nonlinear optimization for minimizing the simulation error (Eq.~\eqref{eqSE}) may be conducted afterwards \cite{Spiridonakos2015}. However, this is not used in the current paper due to the fact that LARS allows one to detect the appropriate NARX terms, therefore the models estimated by ordinary least-squares appear sufficiently accurate.

			 If an appropriate NARX model is not obtained, \ie the initial assumption that the full NARX model includes an appropriate candidate is not satisfied, the process is re-started from Step 1.1 (choice of model class), when different options for the full NARX model should be considered. For instance, one may use more complex models with larger time lags, different basis functions, \etc
		\end{itemize}
		\item Phase 2: Representation of the NARX coefficients by means of PCEs using the sparse adaptive PCE scheme which is based on LARS (see section \ref{sec2.3}). The NARX coefficients obtained from Phase 1 are used for training the PC expansion.
\end{itemize}

\begin{figure}[ht!]
	\centering
	\includegraphics[width=0.65\linewidth]{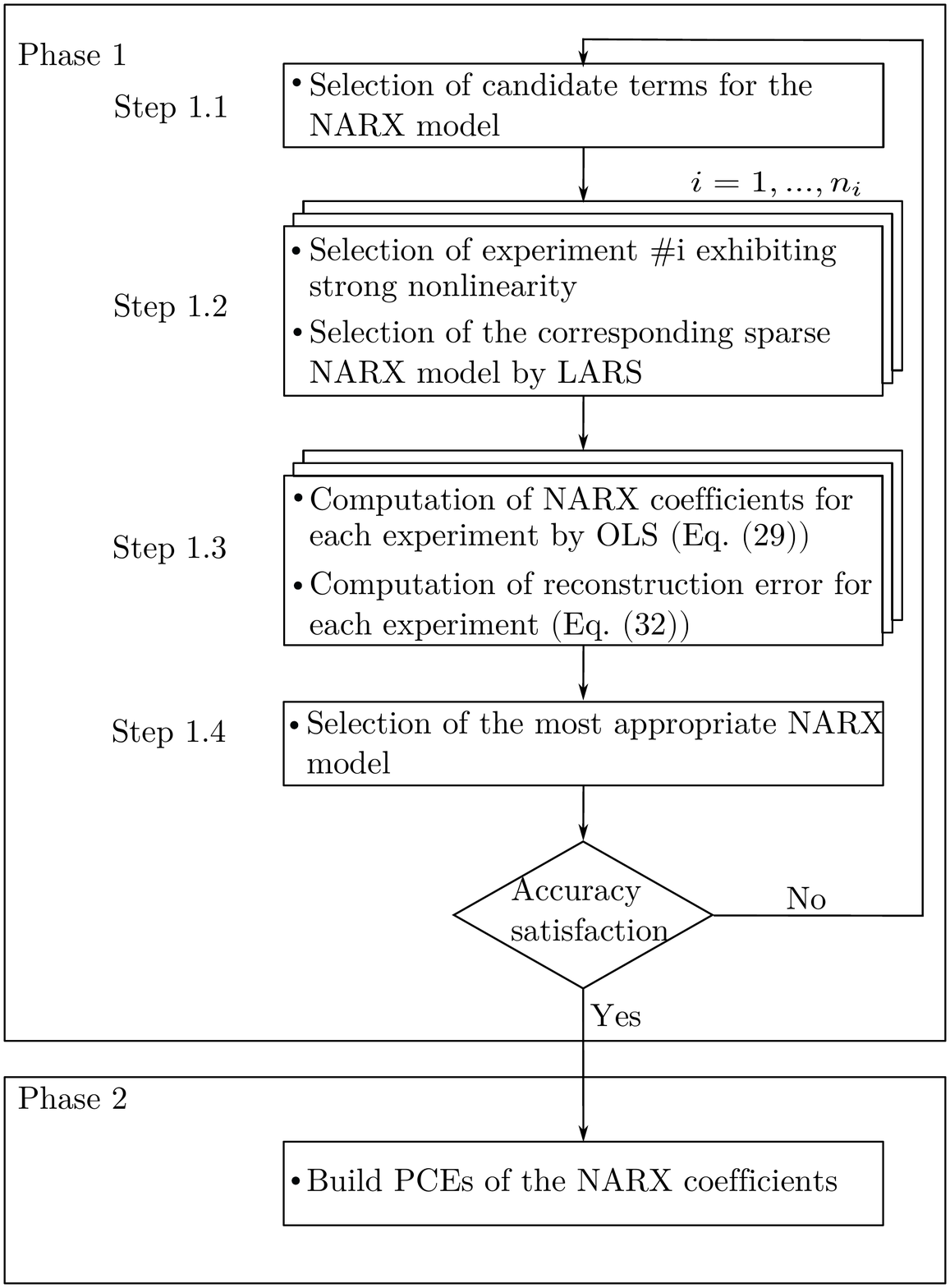}
	\caption{Computation of the LARS-based PC-NARX model}
	\label{fig:flowchartPCNARX}
\end{figure}

For the sake of clarity, the above procedure for computing a PC-NARX model is summarized by the flowchart in \figref{fig:flowchartPCNARX}.

\subsection{Use of the surrogate model for prediction}

The PC-NARX surrogate model can be used for the prediction\footnote{In what follows, the term ``prediction" is employed to refer to the NARX model's so-called ``simulation mode" as addressed in signal processing literature, which stands for the estimation of the response relying only on its initial condition and feedback of the excitation. The term ``prediction" is used however because it is the standard wording in the surrogate modelling community.} of the response to a set of input parameters $\vexi'$. Given the excitation $x(t,\vexi_{x}')$ and the initial conditions of the response $y(t=1,\vexi') = y_{\ve{0}}$, the output time history of the system can be recursively obtained as follows:
\begin{equation}
	\hat{y}(t,\vexi') = \sum\limits_{i=1}^{n_g} \sum\limits_{j=1}^{n_{\psi}} \vartheta_{i,j} \psi_j(\vexi') \, g_i( \hat{\ve{z}}(t,\vexi') ) , \, t = 2 \enum T, 
	\label{eq4.4.1}
\end{equation}
in which
\begin{equation}
	\hat{\ve{z}}(t,\vexi') = \prt{x(t,\vexi_x^{'}) \enum x(t-n_x,\vexi_x^{'}), \hat{y}(t-1,\vexi') \enum \hat{y}(t-n_y,\vexi')}\tr.
\end{equation}

Currently, no close-form formulation for computing the time-dependent statistics of the output quantity is available as opposed to time-frozen PCEs. However, the evolutionary response statistics can be straightforwardly obtained by means of Monte Carlo simulation using the PC-NARX model.

\subsection{Validation of the surrogate model}
The PC-NARX model is computed using an ED of limited size. The validation process is conducted with a validation set of large size which is independent of the ED.
A large number, \eg $n_{val} = 10^4$, of input parameters and excitations is generated. 
One uses the numerical solver to obtain the response time histories sampled at the discrete time instants $t = 1 \enum T$. Then PC-NARX model (Eq.~\eqref{eq4.4.1}) is used to predict the time dependent responses to the excitations and uncertain parameters of the validation set.
The accuracy of the computed PC-NARX model is validated by means of comparing its predictions with the actual responses in terms of
the relative errors and the evolutionary statistics of the response.
For prediction $\# i$, the relative error reads:
\begin{equation}
	\epsilon_{val, i} = \dfrac{ \sum\limits_{t=1}^{T} (y(t, \vexi_i) - \hat{y}(t, \vexi_i))^2 }{\sum\limits_{t=1}^{T} (y(t, \vexi_i) - \bar{y}(t, \vexi_i) )^2} ,
	\label{eq4.4.2}
\end{equation}
where $\hat{y}(t, \vexi_i)$ is the output trajectory predicted by PC-NARX and 
$\bar{y}(t, \vexi_i)$ is the mean value of the actual response time series $y(t,\vexi_i)$.
The above formula is also used to calculate the accuracy of the time dependent statistics (\ie mean, standard deviation) predicted by PC-NARX.
The mean value of the relative errors over $n_{val}$ predictions reads:
\begin{equation}
	\bar{\epsilon}_{val}  = \dfrac{1}{n_{val}} \sum\limits_{i=1}^{n_{val}} \epsilon_{val, i} .
	\label{eq4.4.3}
\end{equation}
The relative error for a quantity $y$, \eg the maximal value of the response (resp. the response at a specified instant) is given by:
\begin{equation}
	\epsilon_{val, y} = \dfrac{ \sum\limits_{i=1}^{n_{val}} (y_{i} - \hat{y}_{i})^2 }{\sum\limits_{i=1}^{n_{val}} (y_{i} - \bar{y} )^2},
\end{equation} 
where $y_i$ is the actual response, $\hat{y}_{i}$ is the prediction by PC-NARX and $\bar{y}$ is the mean value defined by $\bar{y} = \dfrac{1}{n_{val}} \sum\limits_{i=1}^{n_{val}} y_i$.

\section{Numerical applications}
\label{sec5}

The use of LARS-based PC-NARX model is now illustrated with
three nonlinear dynamical systems with increasing complexity, namely a quarter car model subject to a stochastic sinusoidal road profile, a single degree-of-freedom (SDOF) Duffing and a SDOF Bouc-Wen oscillator subject to \emph{stochastic non-stationary} excitation.
In all considered numerical examples, uncertainties arising from the system properties and from the excitation are taken into account. PC-NARX models are computed using a small number of numerical simulations as experimental design. The validation is conducted by comparing their response predictions with the reference values obtained by using Monte Carlo simulation (MCS) on the numerical solvers.

\subsection{Quarter car model}

\begin{figure}[ht!]
	\centering
	\includegraphics[width=0.2\linewidth]{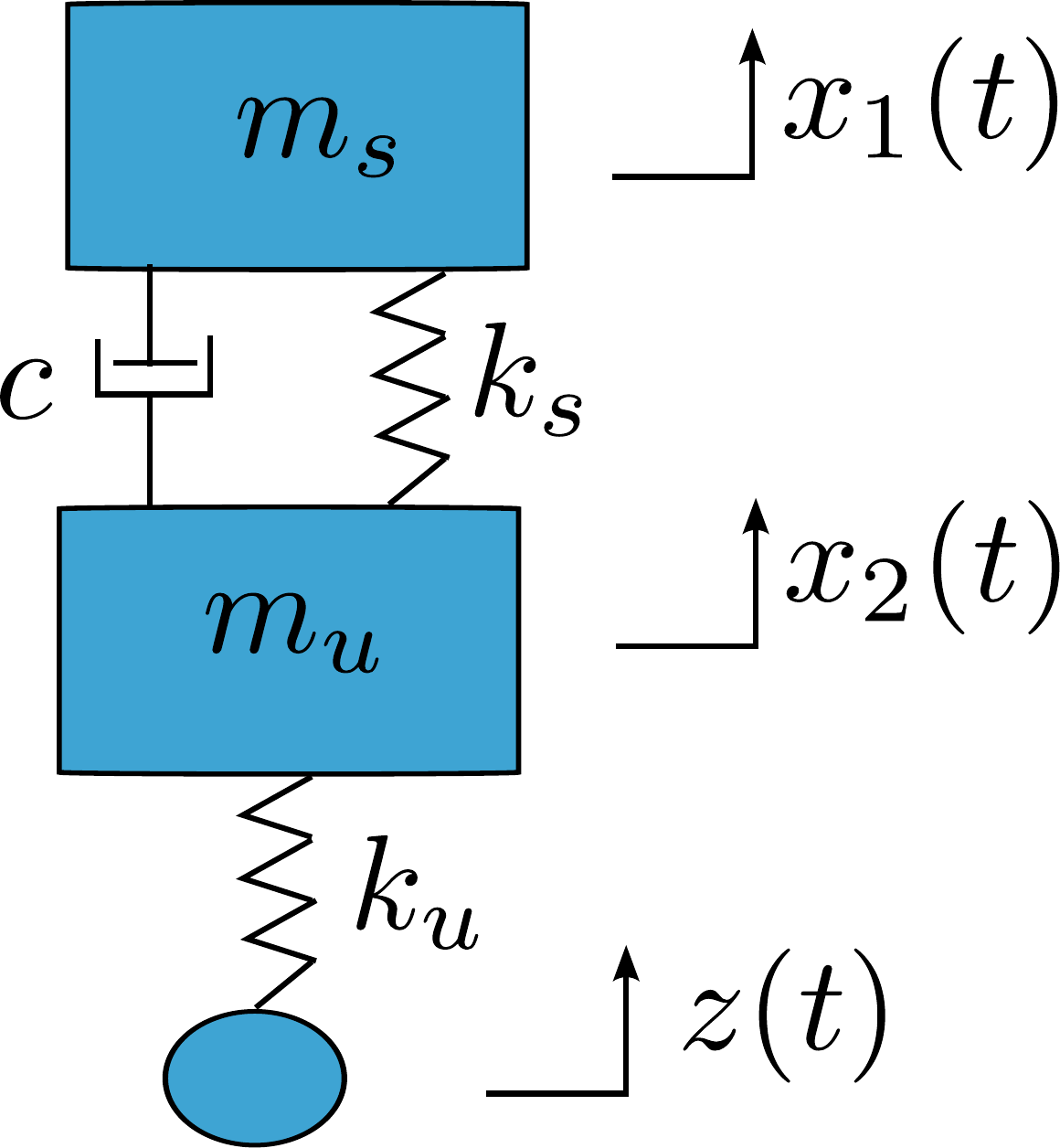}
	\caption{Quarter car model}
	\label{fig5.1.1}
\end{figure}

In the first numerical example, a quarter car model of a vehicle suspension represented by a nonlinear two DOF system \cite{Kewlani2012} (\figref{fig5.1.1}) is considered.
The displacements of the masses are governed by the following system of ordinary differential equations (ODEs):
\begin{equation}
 \left\{
 \begin{array}{l}
  m_s \, \ddot{x}_1(t) = -k_s \, (x_1(t)-x_2(t))^3 -c \, (\dot{x_1}(t) -\dot{x_2}(t)) \\
  m_u \, \ddot{x}_2(t) = k_s \, (x_1(t)-x_2(t))^3 + c \, (\dot{x_1}(t) -\dot{x_2}(t)) + k_u \, (z(t) -x_2)
  \end{array}
 \right.	
 \label{eq5.1.1}
\end{equation}
in which the sprung mass $m_s$ and the unsprung mass $m_u$ are connected by a nonlinear spring of stiffness $k_s$ and a linear damper of damping coefficient $c$. The forcing function $z(t)$ is applied to $m_u$ through a linear spring of stiffness $k_u$. $x_1(t)$ and $x_2(t)$ are the displacements of $m_s$ and $m_u$ respectively. A sinusoidal function road profile with amplitude $A$ and frequency $\omega$ is considered:
\begin{equation}
	z(t) = A \, \sin(\omega \, t).
\end{equation}
%
The parameters of the quarter car model and of the excitation $\vexi = \acc{k_s,\, k_u,\, m_s, \, m_u, \, c, \, A, \, \omega}$ are modelled by independent random variables with marginal distributions given in Table~\ref{tab5.1.1}. Note that the mean value of the parameters are the deterministic values specified in \citet{Kewlani2012}. In addition, Gaussian distributions are used as in \citet{Kewlani2012}, although it would be more appropriate to use \eg lognormal variables to ensure the positivity of mass and stiffness parameters. 
\citet{Kewlani2012} addressed this numerical example with the multi-element PCE approach.

\begin{table}[!ht]
\caption{Parameters of the quarter car model and of the road excitation}
\centering
\begin{tabular}{ccc}
\hline
Parameter & Distribution & Mean \& Standard deviation  \\
\hline
$k_s$ (N/m$^3$) & Gaussian & $(2000,\, 200)$ \\
$k_u$ (N/m) & Gaussian & $(2000,\, 200)$ \\
$m_s$ (kg) & Gaussian & $(20,\,2)$ \\
$m_u$ (kg) & Gaussian & $(40,\,4)$ \\
$c$ (N\,s /m)& Gaussian & $(600,\, 60)$\\
\hhline{===}
Parameter & Distribution & Support  \\
\hline
$A$ (m)& Uniform & $[0.09,\, 0.11]$ \\
$\omega$ (rad/s)& Uniform & $[1.8 \, \pi, \, 2.2 \, \pi ]$ \\
\hline
\end{tabular}
\label{tab5.1.1}
\end{table}

We now aim at building the metamodel for representing the displacement time histories $x_1(t)$ of the sprung mass $m_s$. For this purpose, $N = 100$ analyses of the system are conducted with 100 samples of the uncertain input parameters generated by Latin hypercube sampling. The system of ODEs are solved by means of the Matlab solver \textit{ode45} (explicit Runge-Kutta method with relative error tolerance $1 \, \times 10^{-3}$) for the total duration $T=30$~s and the time step $\di t=0.01$~s. In the first place, a
NARX model structure is chosen, in which the model terms are polynomial functions of past values of the output and excitation $g_i(t) = x_1^l(t-j) \, z^m(t-k) $ with $l+m \leq 3$, $0 \leq l \leq 3$, $0 \leq m \leq 1$, $j= 1\enum 4$, $k = 0\enum 4$. The specified full NARX model contains 86 terms. It is worth noticing that the initial choice of the NARX structure was facilitated by the knowledge of the dynamical nonlinear behaviour of the system of interest. For instance, polynomial functions of order up to 3 are used because of the cubic nonlinear behaviour as in Eq.~\eqref{eq5.1.1}. As a rule of thumb, the maximum time lags $n_x=n_y=4$ are chosen equal to twice the number of degrees of freedom of the considered system. 

Next, the candidate NARX models were computed. To this end, we selected the simulations with maximum displacement exceeding a large threshold, \ie $\max(\abs{x_1(t)}) > 1.2$~m and retained 15 experiments. For each selected simulation, LARS was applied to the initial full NARX model to detect the most relevant NARX terms constituting a candidate NARX model. This procedure resulted in 10 different candidate NARX models in total.

For each candidate NARX model, we computed the NARX coefficients for all simulations by ordinary least-squares to minimize the sum of squared errors (Eq.~\eqref{eqOLSsolution}). We then reconstructed all the output time histories with the computed coefficients and calculated the relative errors $\epsilon_k$ of the reconstruction (Eq.~\eqref{eqSE}).

Among the 10 candidates, we selected the most appropriate NARX model which achieves a sufficiently small overall error with the smallest number of terms. The selected model results in a mean relative error $\bar{\epsilon}= 3.56 \times 10^{-4}$ for 100 simulations in the ED and contains 6 terms, namely the constant term, $z(t-4)$, $x_1(t-4)$, $x_1(t-1)$, $x_1^3(t-1)$, $x_1^2(t-4)\, z(t-4) $. LARS proves effective in selecting the appropriate NARX model by retaining only 6 among the 86 candidate terms available to describe the system. 
	 
In the next step, we expanded the 6 NARX coefficients by adaptive sparse PCEs of order $p\leq20$ with maximum interaction rank $r=2$ and truncation parameter $q = 1$. 
The NARX coefficients computed in the previous step were used for training the metamodel.
PCE models that minimize the LOO errors were selected.
This led to LOO errors smaller than $10^{-7}$. The optimal PCE order selected by the adaptive scheme is up to 6. 

For the sake of comparison, we represent the response $x_1(t)$ by means of time-frozen PCEs. For this purpose, adaptive sparse PCEs are used with an ED of size $N =500$. The best PCE with maximum degree $ 1 \leqslant p \leqslant 20 $, maximum interaction rank $r=2$ and truncation parameter $q = 1$ is selected.
The PC-NARX and time-frozen PCEs are used to predict the output time histories for an independent validation set of size $n_{val} = 10^4$ which is pre-computed by the numerical solver. The accuracy of the two PCE approaches are compared in the following.

\begin{figure}[ht!]
	\centering
\subfigure[$t=5$~s]
	{
	\includegraphics[width=0.45\linewidth]{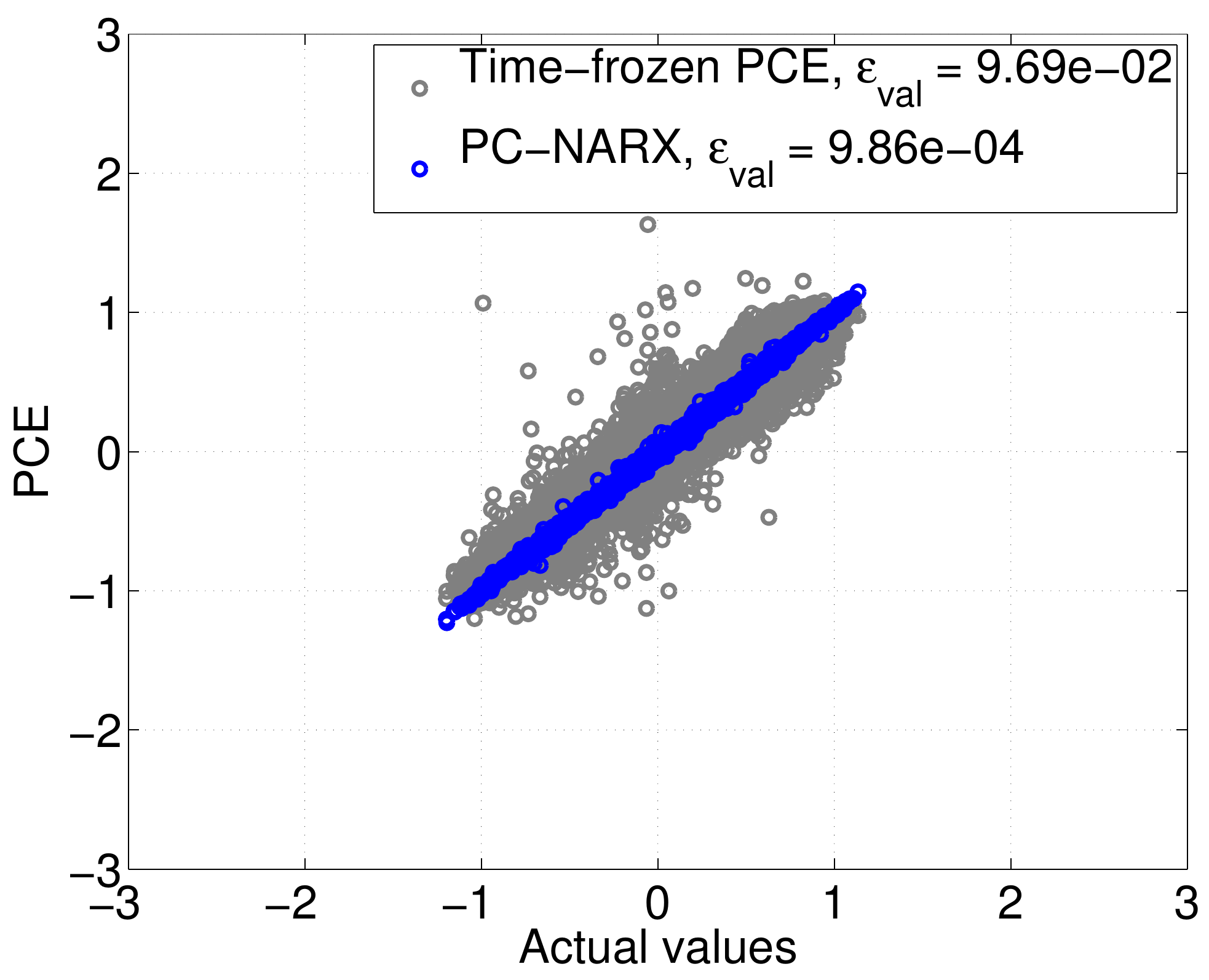}
	\label{fig5.1.4.a}
	}
	\subfigure[$t=5$~s]
	{
	\includegraphics[width=0.45\linewidth]{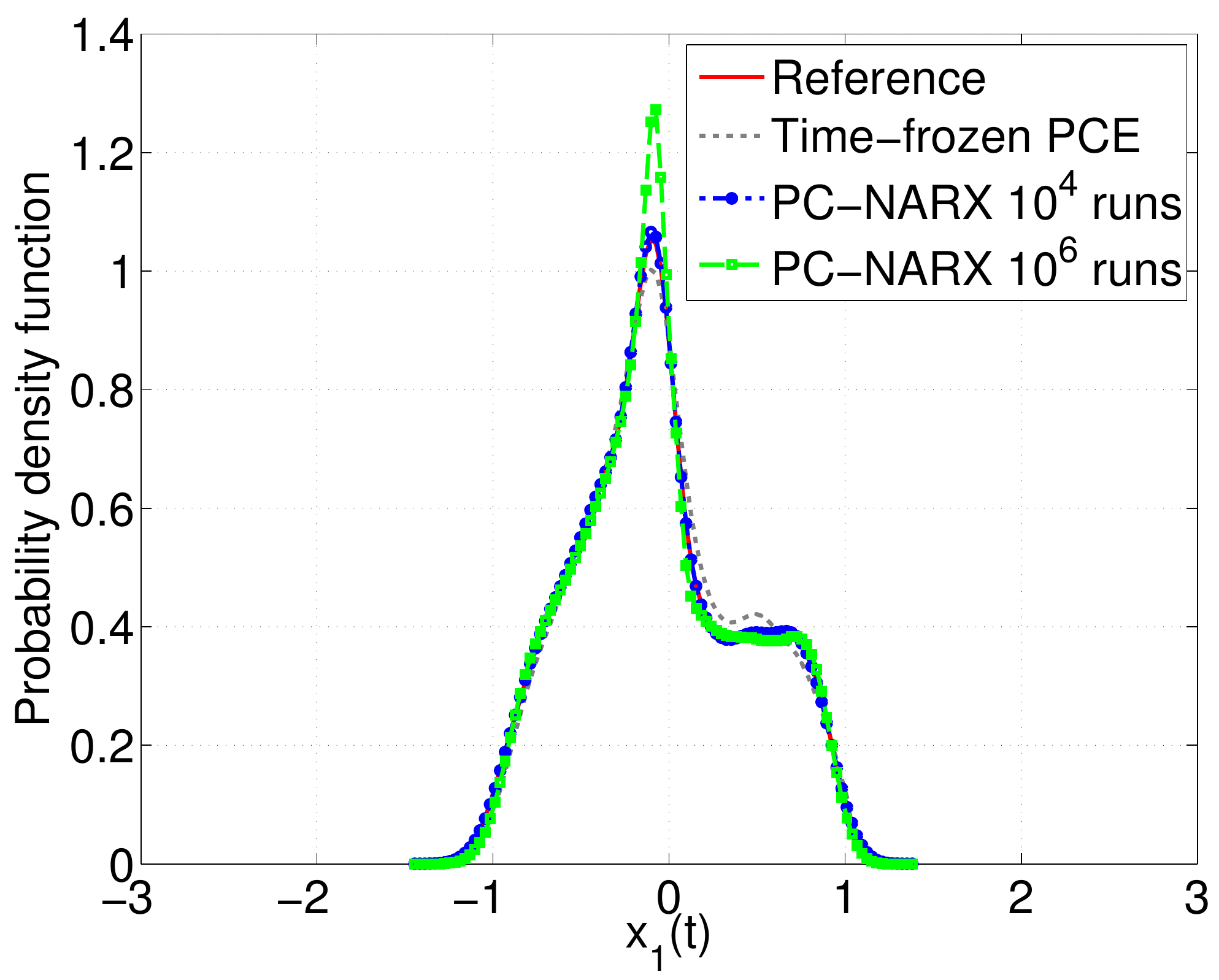}
	\label{fig5.1.4.b}
	}
\subfigure[$t=30$~s]
	{
	\includegraphics[width=0.45\linewidth]{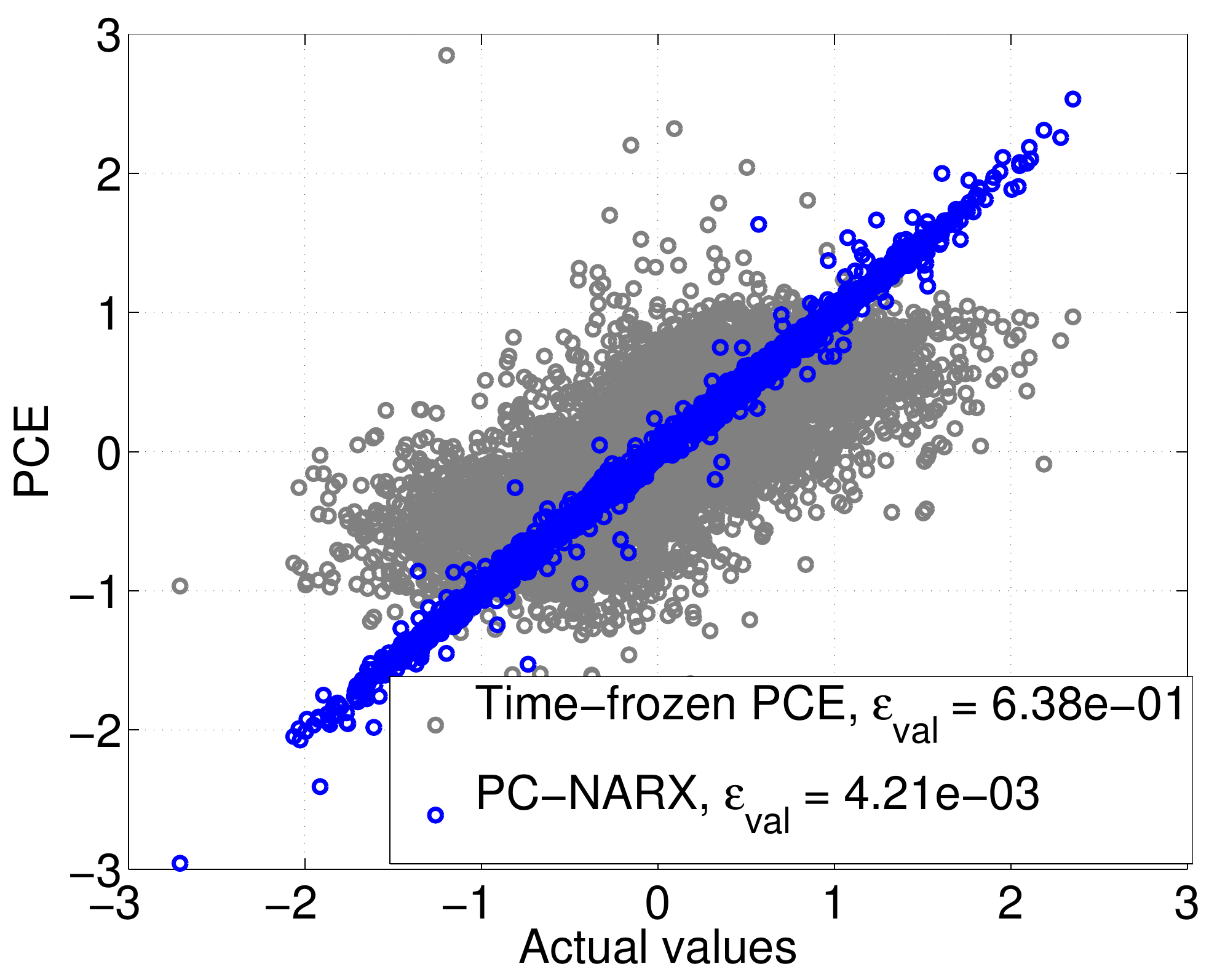}
	\label{fig5.1.4.c}
	}
	\subfigure[$t=30$~s]
	{
	\includegraphics[width=0.45\linewidth]{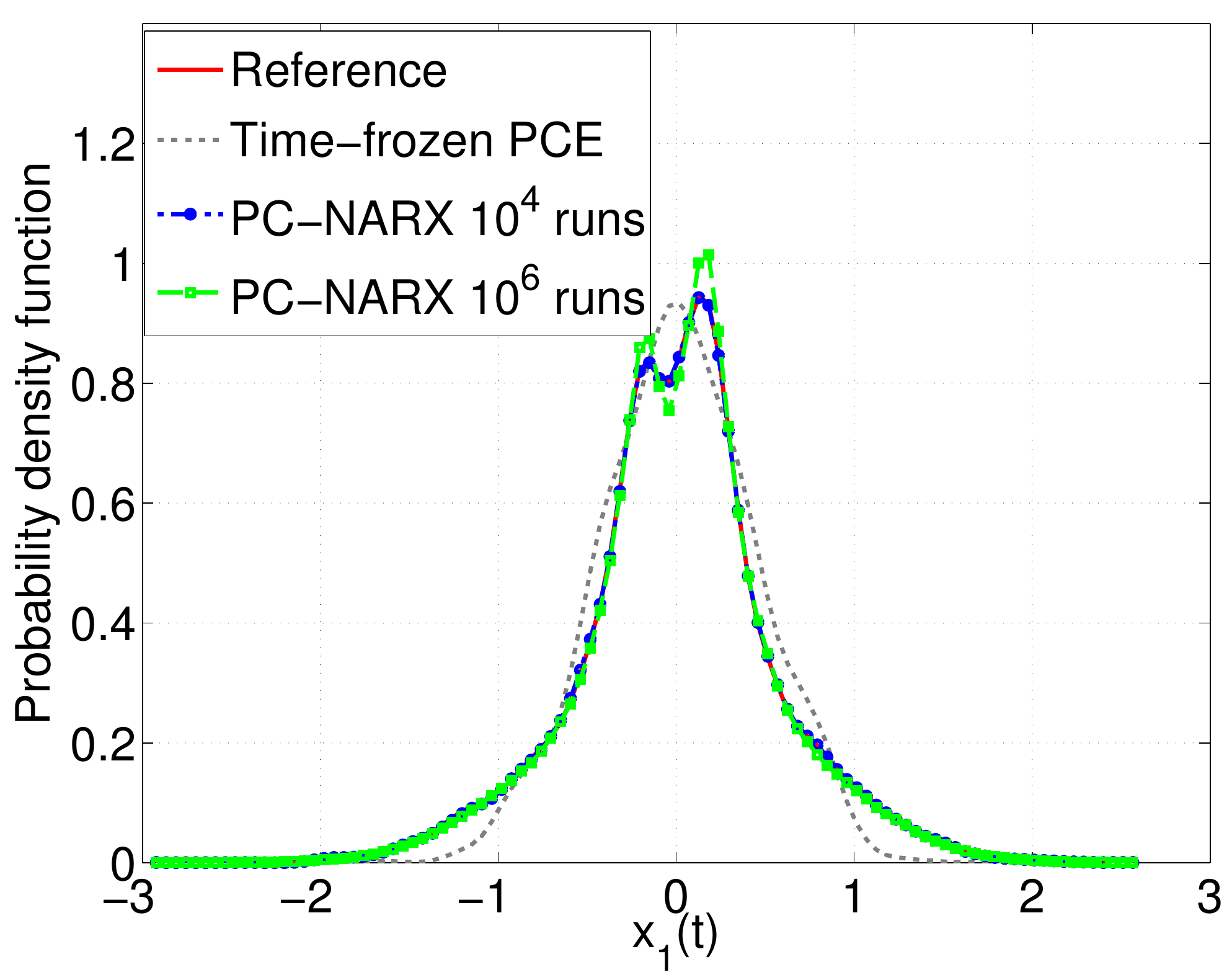}
	\label{fig5.1.4.d}
	}
	\caption{Quarter car model -- Instantaneous displacements: comparison of the two approaches.}
	\label{fig5.1.4}
\end{figure}

We compare the predictions of the mass displacement at an early time instant $t=5$~s and a late instant $t=30$~s (\figref{fig5.1.4}).
The instantaneous responses $x_1(t)$ predicted by two PCE approaches are plotted in \figref{fig5.1.4.a} and \figref{fig5.1.4.c} versus the actual responses obtained by numerically solving
the considered system of ODEs. 
One observes that PC-NARX outperforms time-frozen PCEs. At $t=30$~s, PC-NARX is
still capable of providing highly accurate predictions ($\epsilon_{val}=4.21 \times 10^{-3}$) whereas time-frozen PCEs show large inaccuracies already from $t=5$~s.
In terms of statistics, the probability density functions (PDFs) of the instantaneous responses are depicted in \figref{fig5.1.4.b} and \figref{fig5.1.4.d}.
The reference PDF is obtained from MCS using the validation set of size $n_{val}=10^4$. 
One notices that the instantaneous PDFs computed by time-frozen PCEs do not differ significantly from the reference function, whereas the predictions are actually of poor quality. Indeed, it is always possible that two sets of samples provide identical PDFs while their pair-wise comparison shows large discrepancies.
This example shows a problem that is overlooked in the literature on surrogate models for dynamical systems, when conclusions are commonly drawn only based on the statistics (for instance, PDFs, mean and standard deviation) of the predicted quantities. This might be misleading for judging the accuracy of the metamodel under investigation in predicting \emph{specific} output values.
The PDFs computed by PC-NARX show perfect agreement with the reference functions, which is obvious because every single PC-NARX-based prediction is 
highly accurate. 

We investigate now the effect of the sample size on the MCS estimates. PC-NARX is used to predict the responses for a different validation set of size $n_{val}=10^6$.
\figref{fig5.1.4.b} and \figref{fig5.1.4.d} show that the PDFs obtained with $10^6$ runs, which are closer to the true PDFs, differ slightly from the PDFs obtained with $10^4$ runs. For instance, the peaks of the $10^6$ runs-based PDFs become marginally higher. It is believed that the PDF obtained with $10^6$ runs of the PC-NARX surrogate is the most accurate, although this cannot be validated by running $10^6$ times the computationally expensive model. In general, it is computational expensive to obtain the true PDFs with MCS on the original numerical model, however it is feasible using PC-NARX model.

\begin{figure}[ht!]
	\centering
	\subfigure[Maximal displacement $\max (\abs{x_1(t)})$]
	{
	\includegraphics[width=0.45\linewidth]{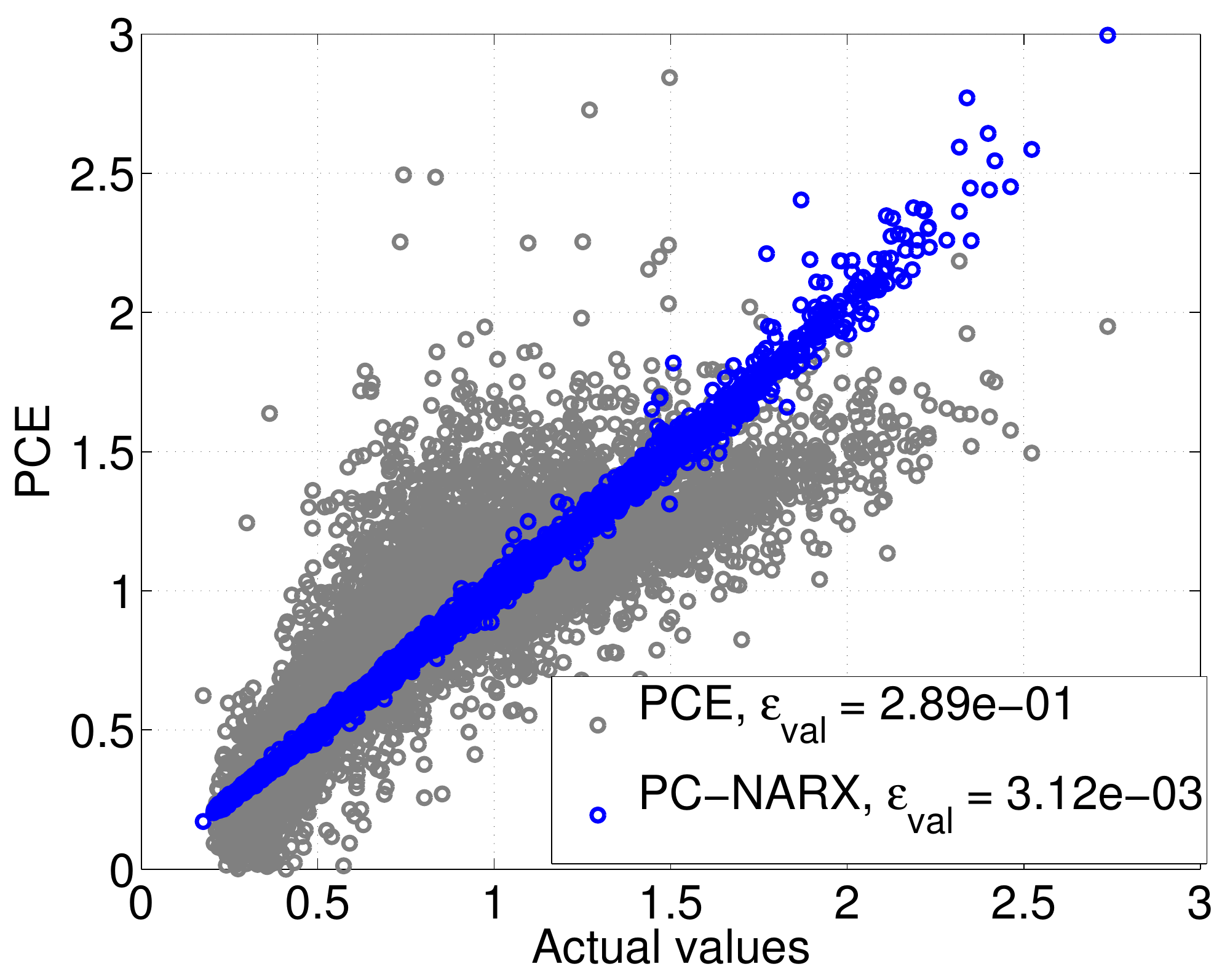}
	}
	\subfigure[PDF of $\max (\abs{x_1(t)})$]
	{
	\includegraphics[width=0.45\linewidth]{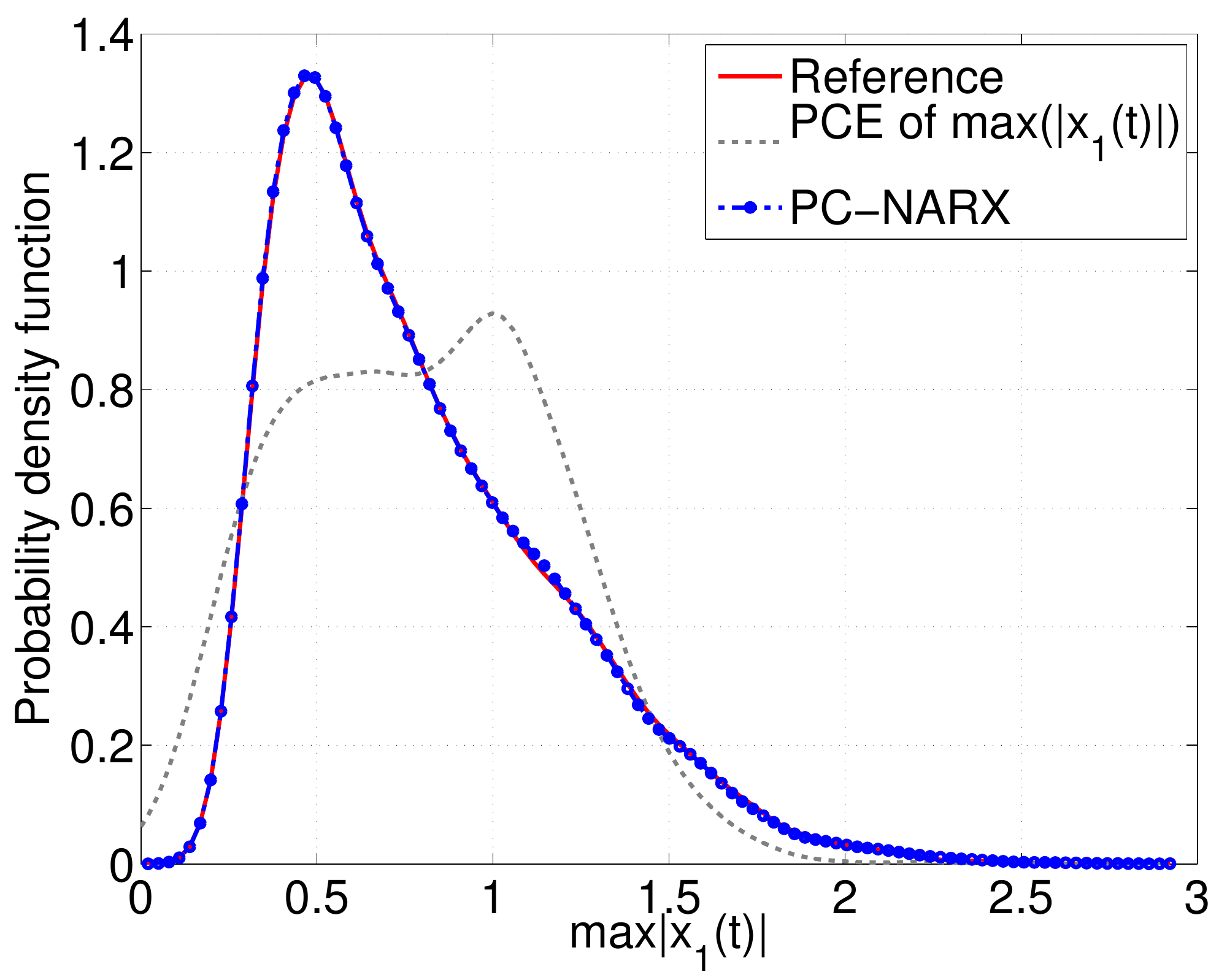}
	\label{fig.ex1.maxpdf}
	}
	\caption{Quarter car model -- Maximal displacement: comparison of the two approaches.}
	\label{fig5.1.5}
\end{figure}

Next, we compare the maximal displacements $\max \prt{\abs{x_1(t)}}$ predicted by the two PCE approaches with the actual values.
Note that the maximal displacements from the ED are retrieved, then used as ED to directly compute the PCE of that quantity.
The same options as in time-frozen PCEs are used. \figref{fig5.1.5} clearly show that PC-NARX outperforms the usual PCE approach in predicting the maximal response. The former provides predictions with great accuracy indicated by the validation error $\epsilon_{val} = 3.12 \times 10^{-3}$,
resulting in a PDF that is consistent with the reference one.
In contrast, the PCE of the maximal displacement gives poor results, as it is expected.
It is already observed that the instantaneous response becomes increasingly complex functions of the input parameters as time evolves.
Consequently, the maximal value, which does not occur at the same time instant for different trajectories, is not a smooth function of the input random variables and cannot be approximated accurately with regular PCEs.
As shown in \figref{fig.ex1.maxpdf}, the PC-NARX technique provides the PDF of this maximum with remarkable accuracy.

\begin{figure}[ht!]
	\centering
	\subfigure[$\vexi = (1897.02,\,	1771.4,\,	22.7,\,	42.0,\,	601.8,\,	0.09,\,	6.00)$]
	{
	\includegraphics[width=0.45\linewidth]{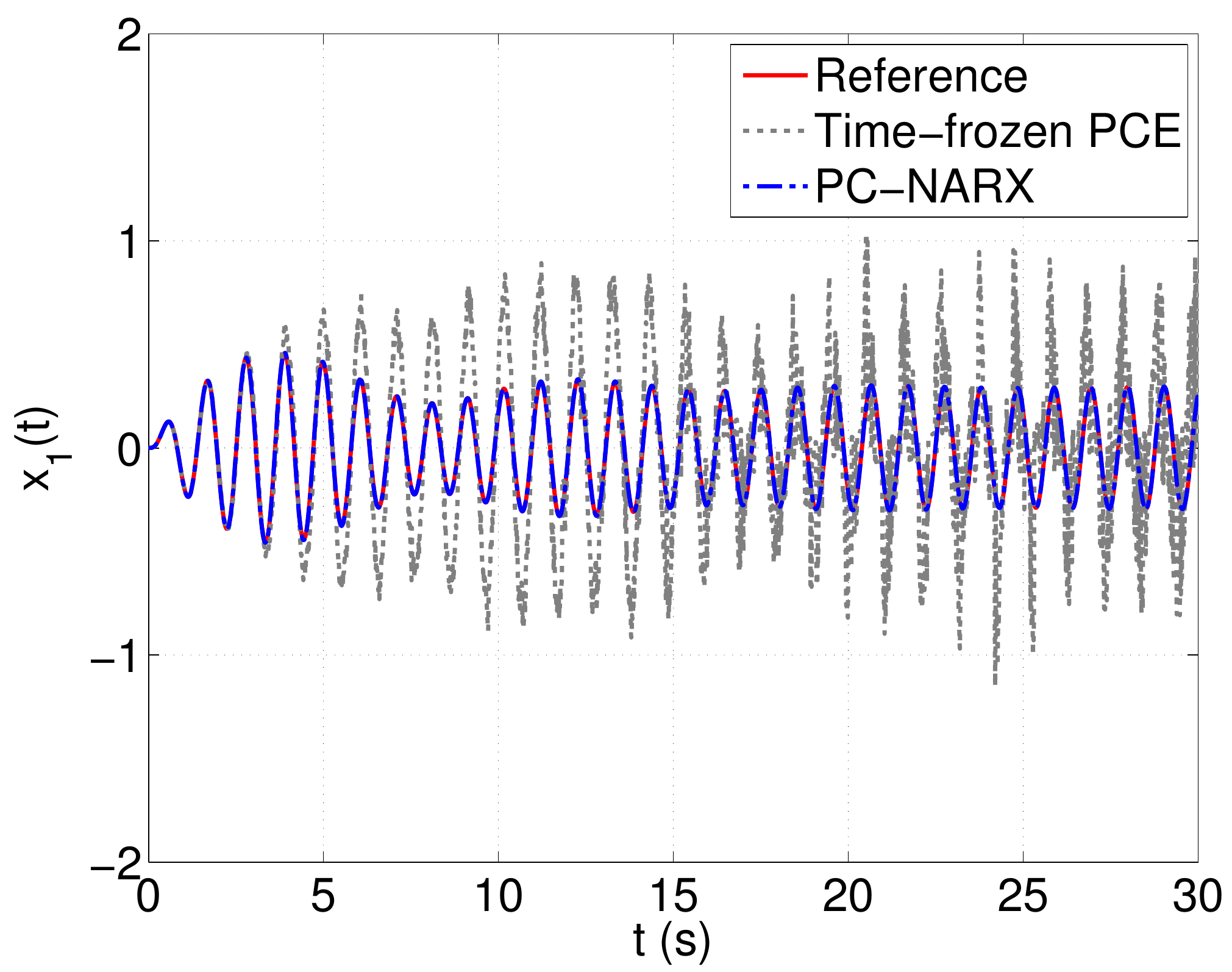}
	}
	\subfigure[$\vexi = (2082.94,\,	2187.10,\,	15.46,\,	40.08,\,	564.26,\,	0.10,\,	6.41)$]
	{
	\includegraphics[width=0.45\linewidth]{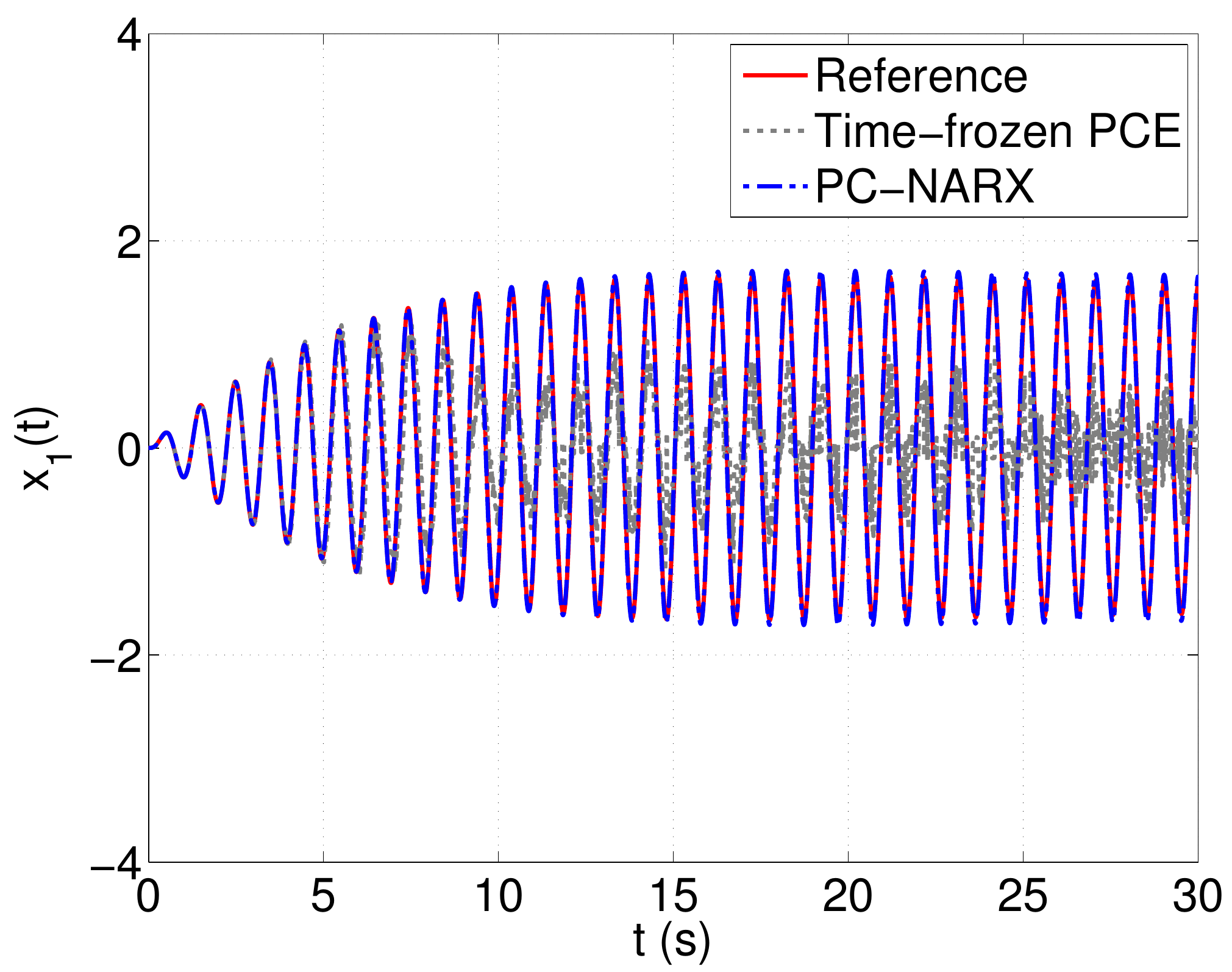}
	}
	\caption{Quarter car model -- Two particular trajectories and their predictions by means of time-frozen PCEs and PC-NARX.}
	\label{fig5.1.6}
\end{figure}

Let us evaluate the overall performance of the two PCE approaches in predicting the entire response time histories.
\figref{fig5.1.6} depicts two specific time dependent response trajectories due to two distinct samples of the uncertain input parameters.
The predictions by the two PCE metamodels are compared with the reference responses obtained by the numerical solver.
It is shown that the accuracy of time-frozen PCEs degenerates relatively quickly as time progresses. Around $t=5$~s, time-frozen PCEs start
showing signs of instability and the predictions become inaccurate. In contrast, PC-NARX provides predictions that are
indistinguishable from the actual responses.
Over $10^4$ validation trajectories, PC-NARX leads to a mean relative error $\bar{\epsilon}_{val} = 0.17 \times 10^{-2}$, and only 5 simulations among them exhibit a relative error $\epsilon_{val, i}$ exceeding $0.1$.

\begin{figure}[ht!]
	\centering
	\subfigure[Mean trajectory]
	{
	\includegraphics[width=0.45\linewidth]{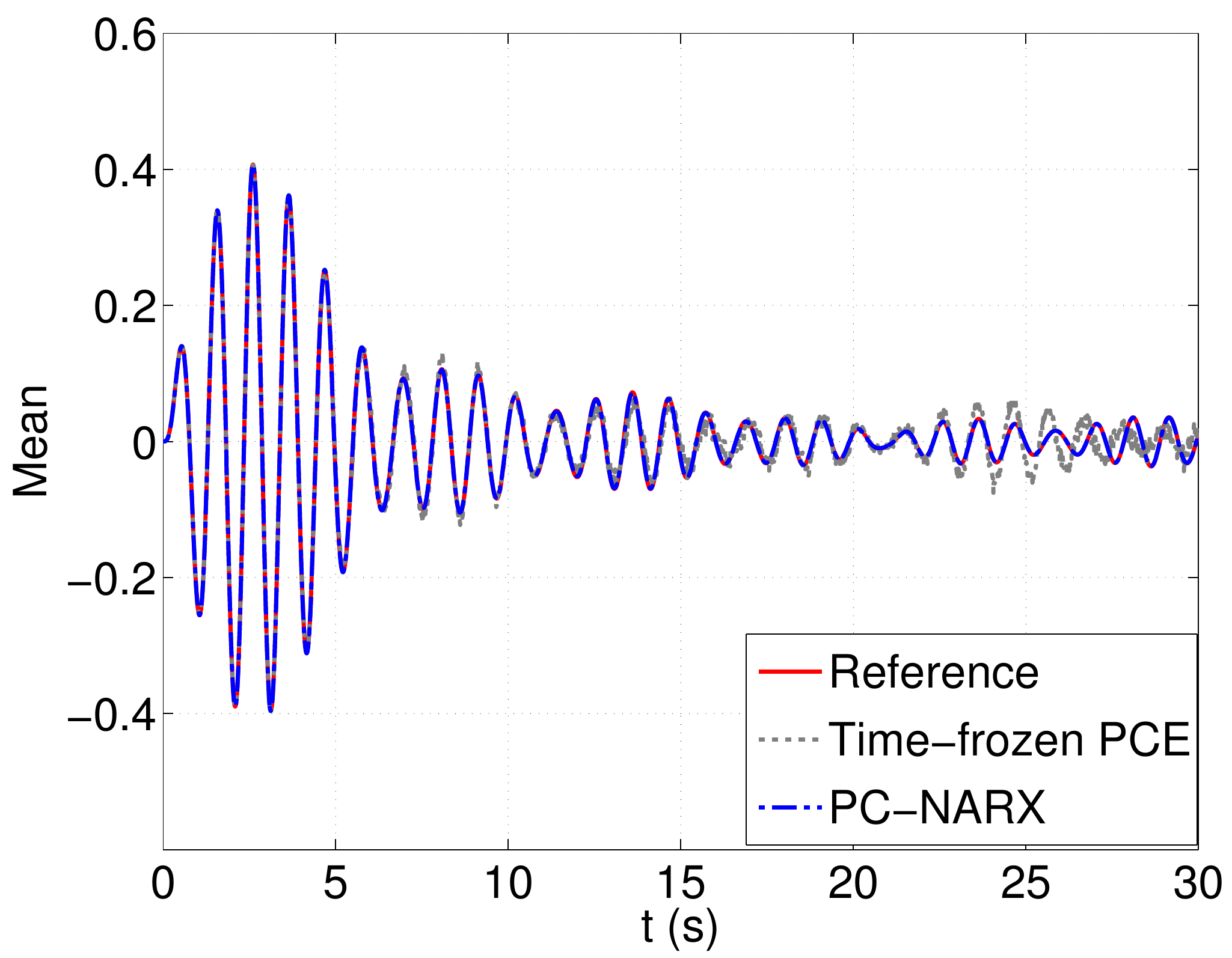}
	}
	\subfigure[Standard deviation]
	{
	\includegraphics[width=0.45\linewidth]{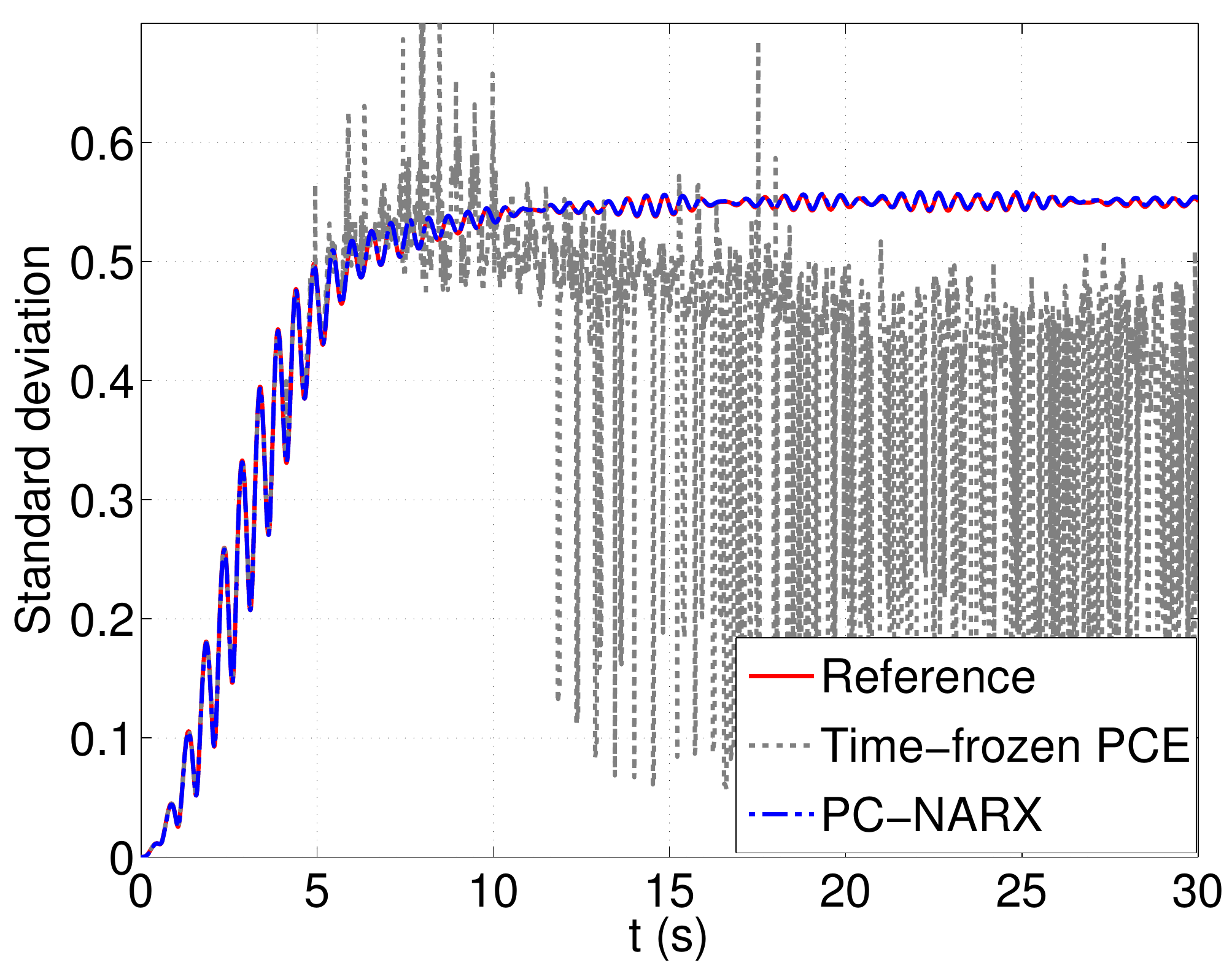}
	}
	\caption{Quarter car model -- Mean and standard deviation of the trajectories: comparison of the two approaches.}
	\label{fig5.1.7}
\end{figure}

The two PCE approaches are now compared in terms of the evolutionary response statistics.
\figref{fig5.1.7} shows that time-frozen PCEs can represent relatively well the mean trajectory, except for the late instants ($t>20$~s) when the discrepancy is noticeable. However, time-frozen PCEs fail to capture the oscillatory behaviour of the standard deviation when the prediction starts deviating significantly from the actual trajectory at $t=5$~s. The improvement in accuracy of PC-NARX is outstanding, in particular because it can represent in detail the oscillatory response statistics. The relative errors are $\epsilon_{val,Mean} = 0.73 \times 10^{-2} $ and $\epsilon_{val, Std} = 0.91 \times 10^{-2}$ for the mean and standard deviation time-histories respectively. 

This numerical case study illustrates the effectiveness of the LARS-based PC-NARX model with respect to the usual PCE approach. It is worth noting that the process was conducted using the uncertainty quantification software UQLab \citep{Marelli2014}, more specifically the polynomial chaos expansion toolbox \citep{UQdoc_09_104}. The next case studies will be more complicated, in particular because the excitations are \emph{non-stationary stochastic processes} with varying intensity and frequency content.

\subsection{Duffing oscillator subject to synthetic ground motions}
\label{sec5.2}
Let us consider a SDOF Duffing oscillator \citep{Kafali2007} subject to stochastic excitation.
Note that the vibration of a Duffing oscillator is a commonly used benchmark in the literature, see \eg \citet{Spiridonakos2015,Orabi1987,Lucor2004,Kougioumtzoglou2009,Chouvion2016} in which an oscillator subject to either periodic or white-noise external force was considered.
The dynamics of the oscillator can be described by the following equation of motion:
\begin{equation}
 \ddot{y}(t) + 2 \, \zeta \, \omega \, \dot{y}(t) +  \omega^2 \, ( y(t) + \epsilon \,  y(t)^3 ) = - x(t) ,
 \label{eq5.2.1}
\end{equation}
in which $y(t)$ is the oscillator displacement, $\zeta$ is the damping ratio, $\omega$ is the fundamental frequency, $\epsilon$ is the parameter governing the nonlinear behaviour and $x(t)$ is the excitation.

Herein, the excitation is generated by the probabilistic ground motion model proposed by \citet{Rezaeian2010}.
The ground motion acceleration is represented as a non-stationary process by means of a modulated filtered white noise process as follows:
\begin{equation}
x(t)= q(t,\ve{\alpha}) \left\{ \frac{1}{\sigma_h{(t)}} \int\limits_{-\infty}^t \!{h[t-\tau,\ve{\lambda}(\tau)] \omega(\tau)}\, \mathrm{d} \tau\right\}.
\label{eq:xt}
\end{equation}
The white-noise process denoted by $\omega(\tau)$ passes a filter $h[t-\tau,\ve{\lambda}(\tau)]$ which is selected as an impulse-response function:
\begin{equation}
\begin{split}
	 h[t-\tau,\ve{\lambda}(\tau)] &= \frac{\omega_f(\tau)}{\sqrt{1-\zeta_f^2(t)}} \mathrm{exp}[-\zeta_f(\tau) \omega_f(\tau) (t-\tau)] \\
	& \times \sin[\omega_f(\tau) \sqrt{1-\zeta_f^2(\tau)} (t-\tau)] \quad \text{for} \quad \tau \leq t ,\\
	 h[t-\tau,\ve{\lambda}(\tau)] &=0 \quad \text{for} \quad \tau > t ,
\end{split}
\end{equation}
where $\ve{\lambda}(\tau)=\left(\omega_f(\tau),\zeta_f(\tau) \right)$ is the vector of time-varying parameters of the filter $h$. $\omega_f(\tau)$ and $\zeta_f(\tau)$ are respectively the filter's frequency and bandwidth at time instant $\tau$. They represent the evolving predominant frequency and bandwidth of the ground motion. A linear model is assumed for $\omega_f(\tau)$ and $\zeta_f(\tau)$ is constant during the entire signal duration:
\begin{equation}
	\omega_f(\tau) =\omega_{mid} + \omega'(\tau- t_{mid}) \quad \mathrm{and} \quad \zeta_f(\tau)=\zeta_f ,
\end{equation}
in which $t_{mid}$ is the instant at which 45\% of the expected Arias intensity $I_a$ is reached, $\omega_{mid}$ is the filter frequency at instant $t_{mid}$ and $\omega'$  is the slope of linear evolution of $\omega_f(\tau)$. After being normalized by the standard deviation $\sigma_h{(t)}$, the integral in Eq.~\eqref{eq:xt} becomes a unit variance process with time-varying frequency and constant bandwidth, which represents the spectral non-stationarity of the ground motion.

The non-stationarity in intensity is then captured by the modulation function $q(t,\ve{\alpha})$. This time-modulating function determines the shape, intensity and duration of the motion as follows:
\begin{equation}
	q(t,\ve{\alpha})=\alpha_1 t^{\alpha_2-1} \mathrm{exp}(-\alpha_3 \,t).
\end{equation}
The vector of parameters $\ve{\alpha}= \prt{\alpha_1, \,\alpha_2,\, \alpha_3}$ is directly related to the physical characteristics of the ground motion, namely the expected Arias intensity $I_a$, the time interval $D_{5-95}$ between the instants at which the 5\% and 95\% of $I_a$ are reached and the instant $t_{mid}$.

In the discrete time domain, the synthetic ground motion in Eq.~\eqref{eq:xt} becomes:
\begin{equation}
	\hat{x}(t)=q(t,\ve{\alpha})\sum_{i=1}^n s_i(t,\lambda(t_i)) U_i ,
	\label{eq3.5}
\end{equation}
where the standard normal random variable $U_i$ represents an impulse at instant $t_i$ and $s_i(t,\lambda(t_i))$ is given by:
\begin{align}
	s_i(t,\lambda(t_i))
	&= \frac{h[t-t_i,\lambda(t_i)]}{\sqrt{\sum_{j=1}^k h^2[t-t_j,\lambda(t_j)]}}  ~ \text{for} ~ t_i < t_k, \, t_k\leq t < t_{k+1}  ,\\
	&=0  \quad \text{for} \quad t\leq t_i .
	 \notag
\end{align}

Herein, the parameters $\zeta$ and $\omega$ of the SDOF oscillator are considered deterministic with values $\zeta = 0.02$ and $\omega = 5.97$~rad/s. The uncertain input vector contains parameters of the oscillator and parameters representing the main intensity and frequency features of the ground motion model 
$\vexi = \prt{\epsilon, I_a, D_{5-95}, t_{mid}, \omega_{mid}, \omega', \zeta_f}$.
Table~\ref{tab5.2.2} represents the probabilistic distributions associated with
the uncertain parameters.
The six parameters describing the ground motion are considered dependent with a Nataf distribution (a.k.a Gaussian copula) \citep{DK86b,Lebrun2009a}.
The correlation matrix is given in Table~\ref{tab5.2.3}.

\begin{table}[!ht]
\caption{Marginal distributions of the stochastic ground motion model (after \citet{Rezaeian2010}) and of the Duffing oscillator's non linearity $\epsilon$.}
\centering
\begin{tabular}{ccccc}
\hline
Parameter & Distribution & Support & Mean & Standard deviation \\
\hline
$I_a$ (s.g) & Lognormal & (0, +$\infty$) & 0.0468 & 0.164 \\
$D_{5-95}$ (s) & Beta      & [5, 45]  & 17.3 & 9.31 \\
$t_{mid}$ (s) & Beta & [0.5, 40] & 12.4 & 7.44 \\
$\omega_{mid}$/{2$\pi$} (Hz) & Gamma & (0, +$\infty$) & 5.87 & 3.11 \\
$\omega'$/{2$\pi$} (Hz) & Two-sided exponential&[-2, 0.5] &-0.089 & 0.185 \\
$\zeta_f$ (.) & Beta & [0.02, 1] & 0.213 & 0.143\\
\hline
$\epsilon$ (1/m$^2$) & Uniform & $[90,110]$  & 100 & 5.773 \\
\hline
\end{tabular}
\label{tab5.2.2}
\end{table}

\begin{table}[!ht]
\caption{Correlation matrix of the Nataf distribution of the stochastic ground motion model (after \citet{Rezaeian2010}).}
\centering
\begin{tabular}{ccccccc}
\hline
		& $I_a$ & $D_{5-95}$ & $t_{mid}$ & $\omega_{mid}$ & $\omega'$ & $\zeta$ \\
\hline
$I_a$ & 1 & -0.36 & 0.01 & -0.15 & 0.13 & -0.01 \\
$D_{5-95}$  & -0.36 & 1 & 0.67 & -0.13 & -0.16 & -0.2 \\
$t_{mid}$ & 0.01 & 0.67 & 1 & -0.28 & -0.2 &-0.22 \\
$\omega_{mid}$ & -0.15 & -0.13 & -0.28 & 1 & -0.2 & 0.28 \\
$\omega'$  & 0.13 & -0.16 & -0.2 & -0.2 & 1 & -0.01 \\
$\zeta$ & -0.01 & -0.2 & -0.22 & 0.28 & -0.01 & 1\\
\hline
\end{tabular}
\label{tab5.2.3}
\end{table}

We now aim at building the metamodel for representing the displacement time histories $y(t)$. For this purpose, $N = 200$ analyses of the system were conducted with the input parameters generated by Latin hypercube sampling. The equation of motion was solved by means of the Matlab solver \textit{ode45} (explicit Runge-Kutta method with relative error tolerance $1 \,\times 10^{-3}$). Note that for the sake of consistency, all the synthetic motions were generated with the total duration $T=30$~s and time step $\di t=0.005$~s. In the first place, a
NARX model structure was chosen in which the model terms are polynomial functions of past values of the output and excitation $g_i(t) = y^l(t-j) \, x^m(t-k) $ with $l+m \leq 3$, $0 \leq l \leq 3$, $0 \leq m \leq 1$, $j= 1,2$, $k = 0, 1,2$. The chosen full NARX model contains 10 terms.

Next, candidate NARX models were computed. To this end, we selected the simulations with maximum displacement exceeding a large threshold, \ie $\max(\abs{y(t)}) > 0.07$~m, leading to 19 selected experiments. For each simulation previously selected, LARS was applied to the initial full NARX model to detect the most relevant NARX terms constituting a candidate NARX model. This procedure resulted in 12 candidates in total. 
The NARX coefficients corresponding to each candidate model are then computed for each simulation in the ED by means of ordinary least squares (Eq.~\eqref{eqOLSsolution}). The responses are reconstructed using the computed coefficients, leading to the errors $\epsilon_k$ (Eq.~\eqref{eqSE}).
The best NARX model achieves a mean relative error $\bar{\epsilon}= 7.4 \, \times 10^{-4}$ over 200 experiments and contains 7 terms, namely the constant term, $x(t-2)$, $x(t)$, $y(t-2)$, $y(t-1)$, $y^2(t-2)$, $y^3(t-1)$. 

In the next step, we represented the NARX coefficients by adaptive sparse PCEs of order up to 20 with maximum interaction rank $r=2$ and truncation parameter $q = 1$. 
The PCEs of the NARX coefficients have LOO errors smaller than $7.34\,\times 10^{-4}$. The optimal PCE selected by the adaptive scheme is of total degree 3.

\begin{figure}[ht!]
	\centering
	\subfigure[First example trajectory $\vexi = (106.30, 0.05,    6.53,    5.47,   55.16,   -1.02,    0.31)$]
		{
		\includegraphics[width=0.45\linewidth]{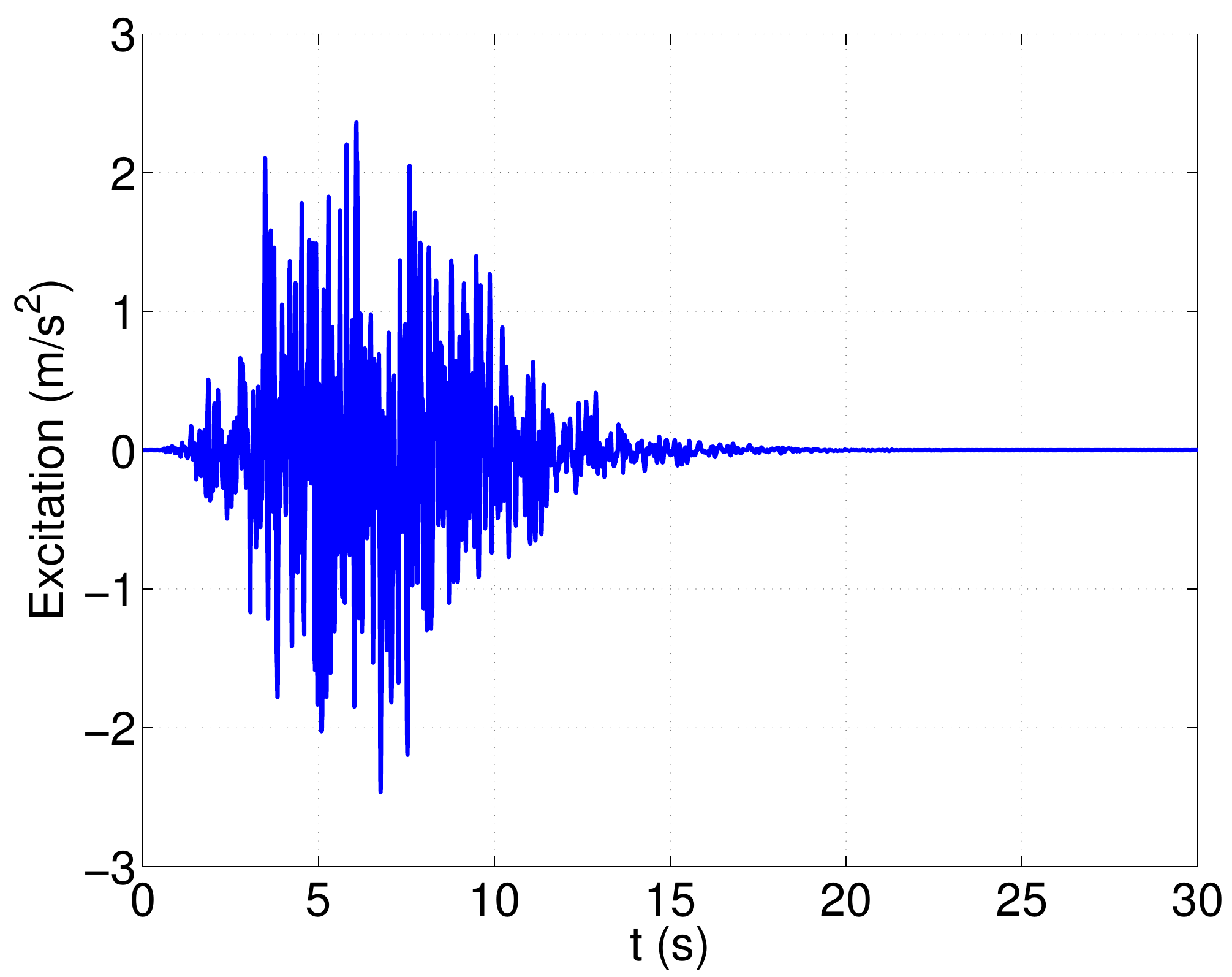}
		}
	\subfigure
				{
				\includegraphics[width=0.45\linewidth]{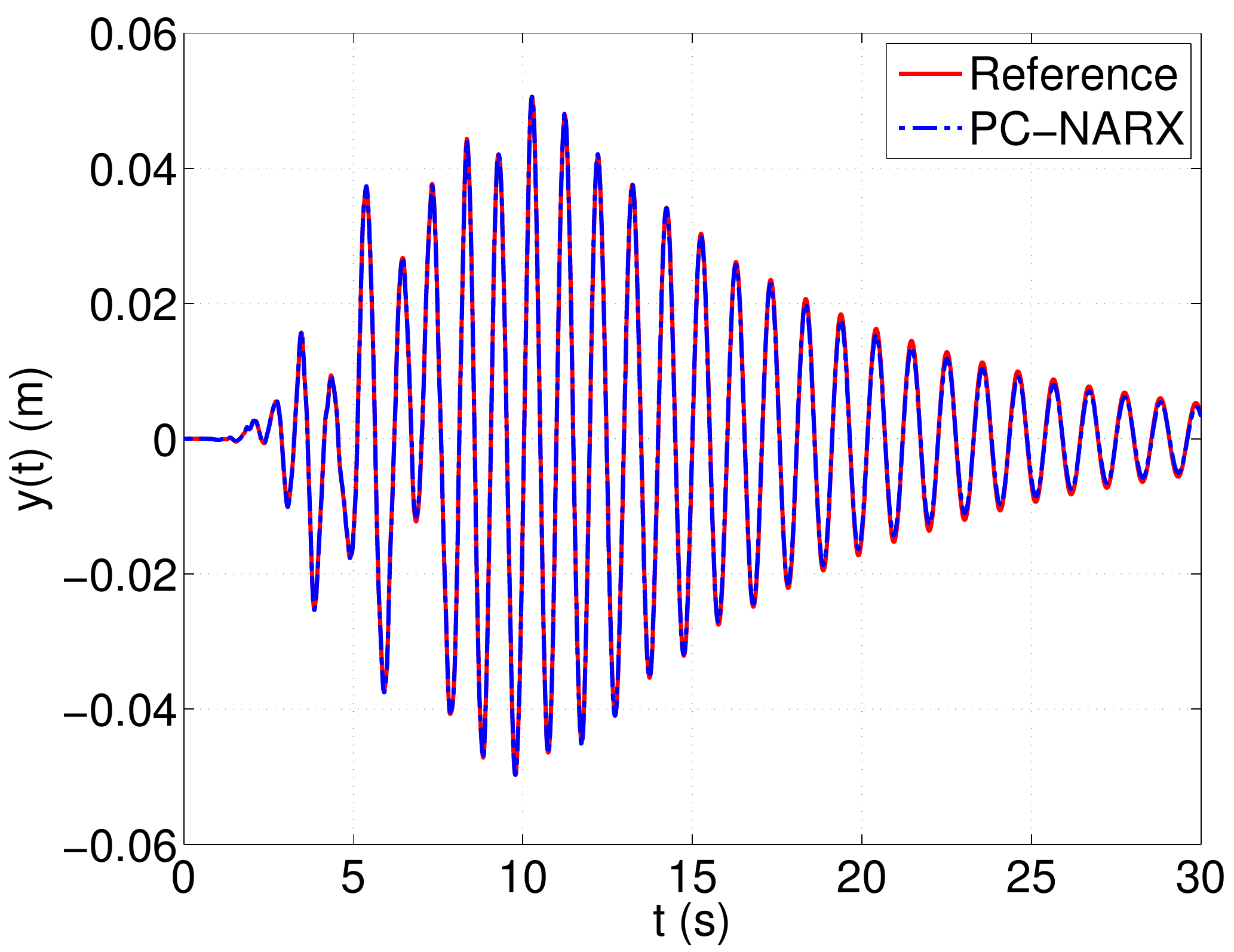}
				}
	\subfigure[Second example trajectory $\vexi = (105.84, 0.05,    9.08,    4.53,   19.76,   -0.11,    0.28)$]
			{
					 \addtocounter{subfigure}{-1}
			\includegraphics[width=0.45\linewidth]{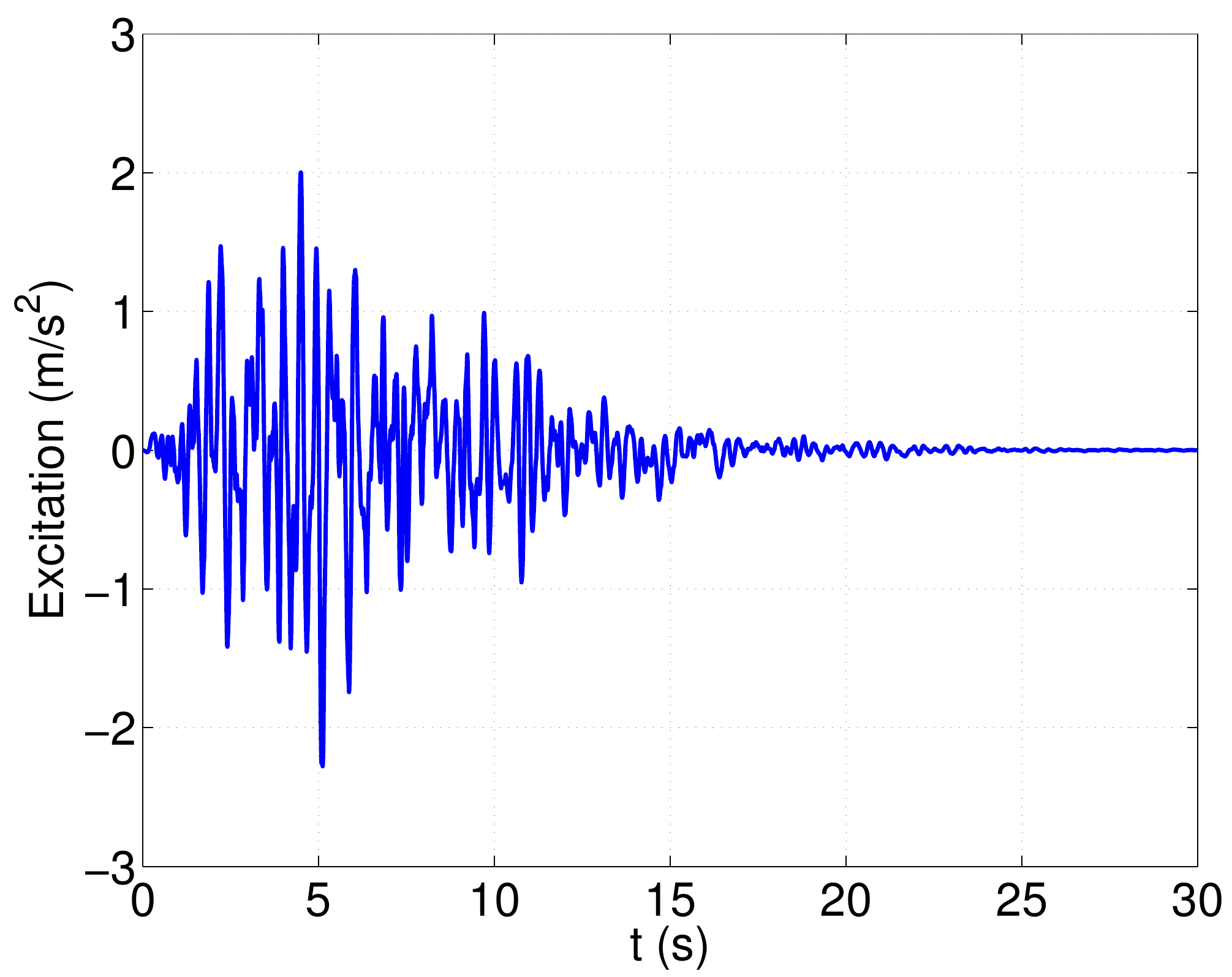}
			}
	\subfigure
		{
		\includegraphics[width=0.45\linewidth]{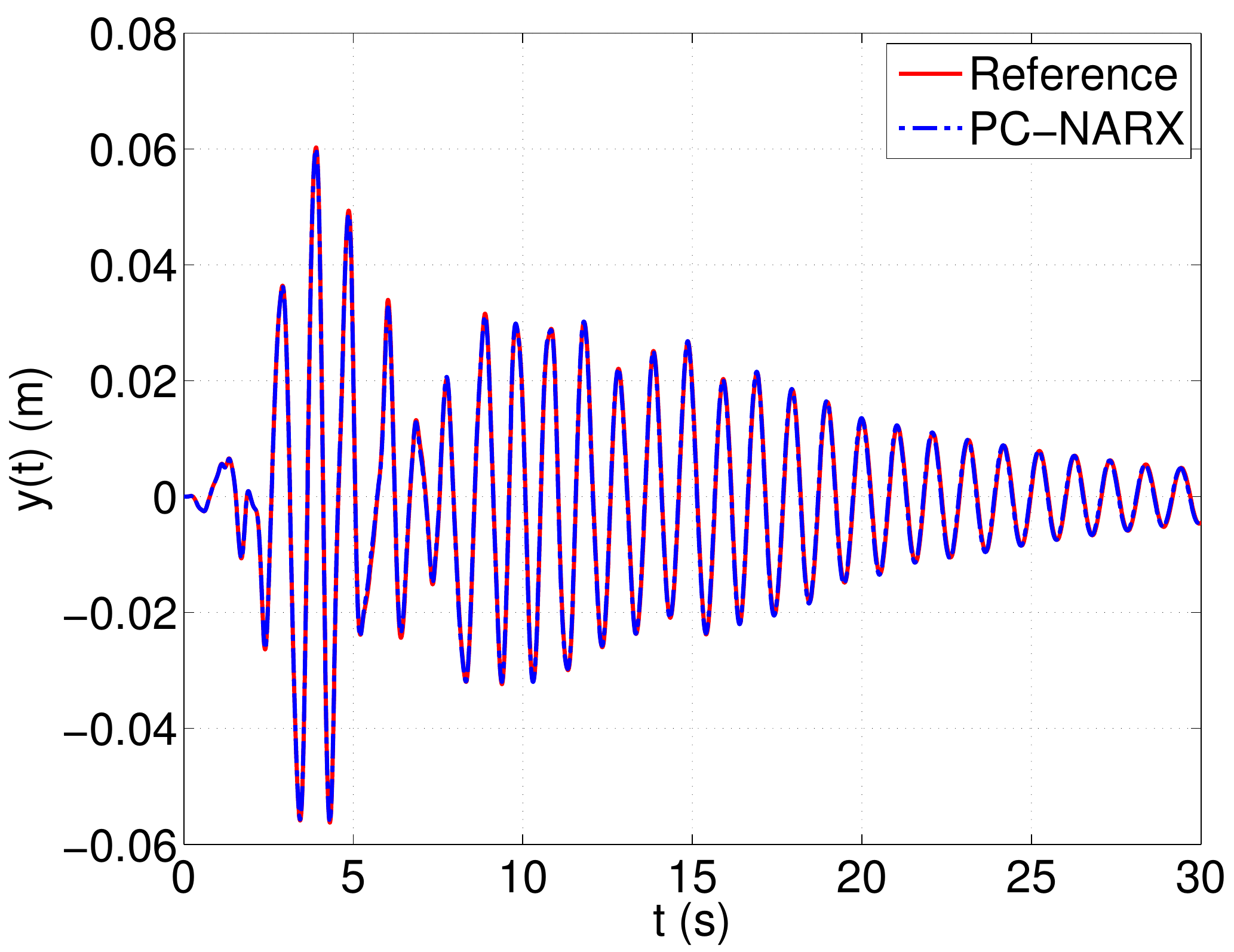}
		}
		\caption{Duffing oscillator -- Two particular excitations, associated response trajectories and their prediction by means of PC-NARX.}
	
	\label{fig5.2.4}
\end{figure}

A validation set of size $n_{val}= 10^4$ was pre-computed by the numerical Matlab solver. The displacements were then predicted by the PC-NARX model. \figref{fig5.2.4} depicts two specific excitations and the corresponding response time histories. PC-NARX provides predictions that are in remarkable agreement with the actual responses. Over $10^4$ validation trajectories, the mean relative error is $\bar{\epsilon}_{val}= 3.53 \times 10^{-2}$. Less than $5\%$ of those simulations exhibit a relative error $\epsilon_{val, i}$ exceeding $0.1$.
Note that predicting the response of a mechanical system subject to \emph{nonstationary excitation} is never an easy task. From our experience, it is of no interest to apply time-frozen PCEs to this type of problems.

\begin{figure}[ht!]
	\centering
	\includegraphics[width=0.45\linewidth]{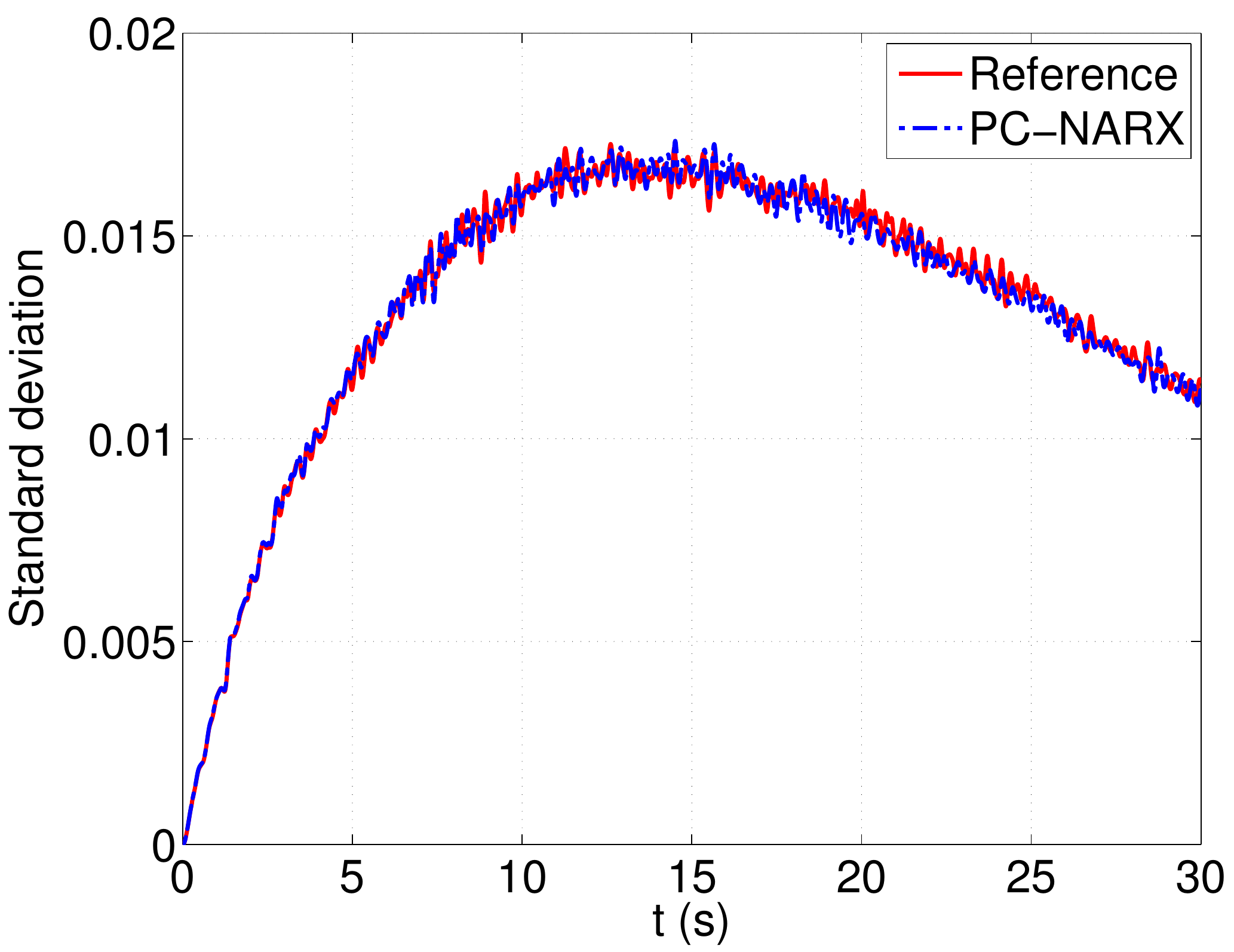}
	\caption{Duffing oscillator -- Standard deviation trajectory of the displacement. }
	\label{fig5.2.5}
\end{figure}

\figref{fig5.2.5} represents the evolutionary standard deviation of the displacement. 
The curve computed by PC-NARX is in excellent agreement with the reference one with a relative error $\epsilon_{val,Std} =  0.5 \times 10^{-2}$. The mean trajectory, which is slightly fluctuating around zero, is not informative therefore not presented herein.

\begin{figure}[ht!]
	\centering
	\subfigure[PCE-based predictions \emph{vs.} actual values]
	{
	\includegraphics[width=0.45\linewidth]{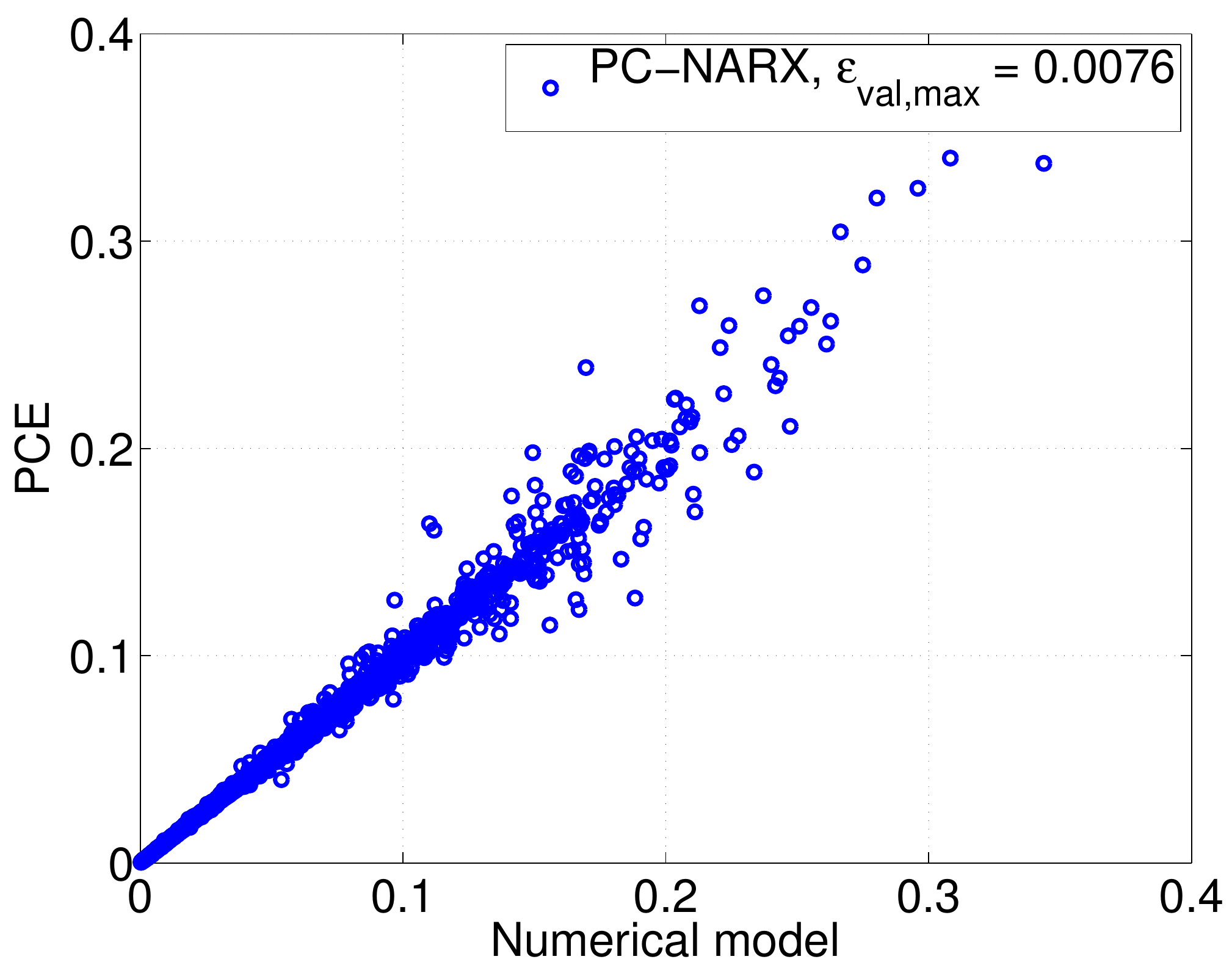}
	}
	\subfigure[Probability density function]
	{
	\includegraphics[width=0.45\linewidth]{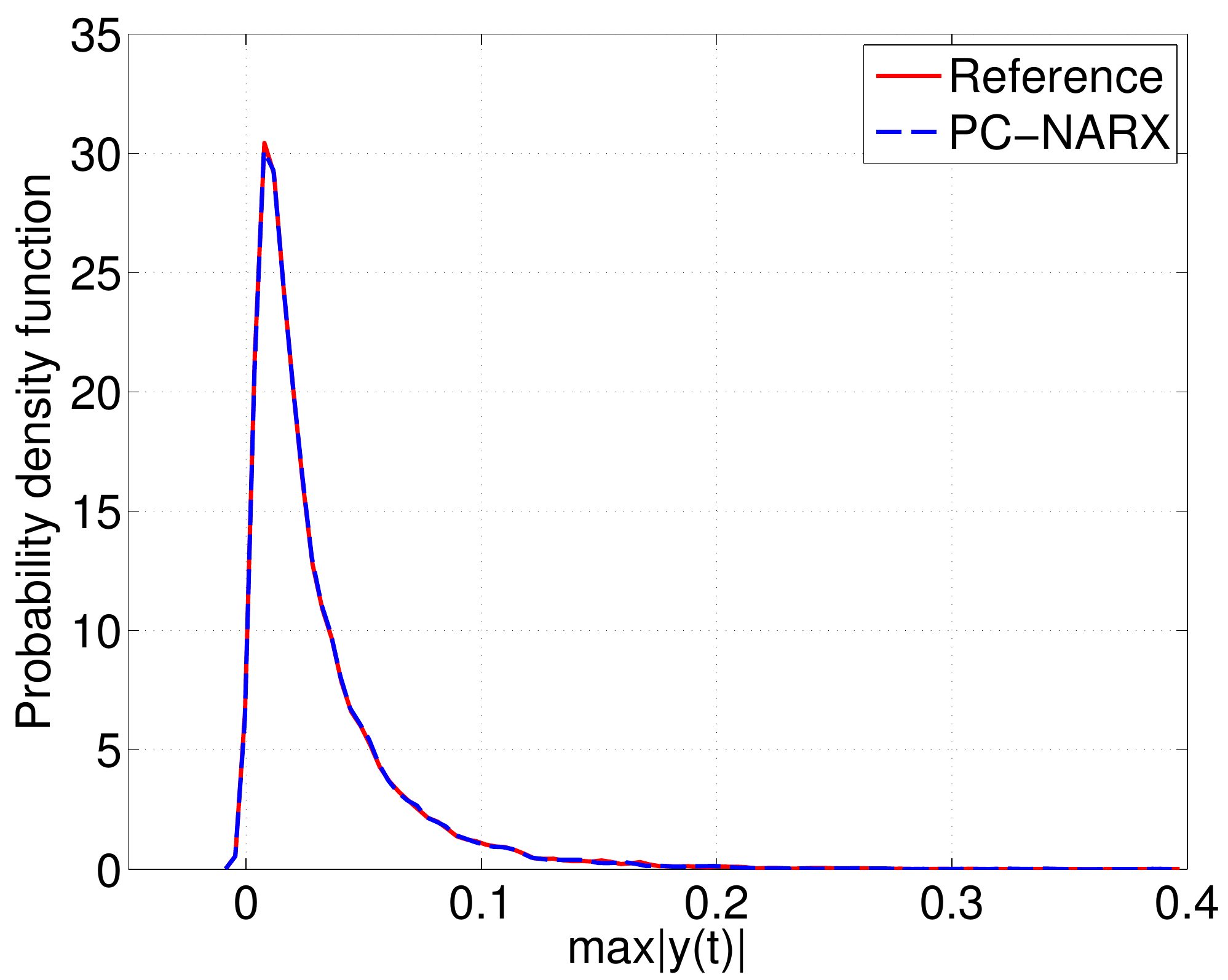}
	}
	\caption{Duffing oscillator -- Maximal displacements. }
	\label{fig5.2.6}
\end{figure}

Finally, we compare the maximal displacements predicted by PC-NARX with the actual values computed by the numerical solver. \figref{fig5.2.6} shows that overall the predictions are quite accurate with a relative error $\epsilon_{val,max} = 0.7 \times 10^{-2}$. The simulations with large discrepancies between the prediction and the actual values correspond to large displacements, \ie they belong to the domain of rare events (the upper tail of the distribution) that was not well represented in the small training set of size 200.

\subsection{Bouc-Wen oscillator subject to synthetic ground motions}

In this application, we consider the SDOF Bouc-Wen oscillator \citep{Kafali2007} subject to stochastic excitation. The equation of motion of the oscillator reads:
\begin{equation}
 \left\{
 \begin{array}{l}
    \ddot{y}(t) + 2 \, \zeta \, \omega \, \dot{y}(t) + \omega^2 (\rho \, y(t) + (1-\rho) \, z(t) ) =  - x(t) \\
    \dot{z}(t) = \gamma \dot{y}(t)  - \alpha \, \abs{\dot{y}(t)} \, \abs{z(t)}^{n-1} z(t) - \beta \, \dot{y}(t) \, \abs{z(t)}^n
 \end{array}
 \right.	
 \label{eq5.3.1}
\end{equation}
in which $\zeta$ is the damping ratio, $\omega$ is the fundamental frequency, $\rho$ is the post- to pre-yield stiffness ratio, $\gamma$, $\alpha$, $\beta$, $n$ are parameters governing the hysteretic loops and the excitation $x(t)$ is a ground motion generated by means of the probabilistic model presented in Section~\ref{sec5.2}. 

Deterministic values are used for the following parameters of the Bouc Wen model: $\zeta = 0.02$, $\rho =0$, $\gamma=1$, $n=1$, $\beta=0$. The remaining parameters are considered independent random variables with associated distributions given in Table~\ref{tab5.3.1}.
The uncertainties in the considered system is therefore characterized by means of vector of uncertain parameters $\vexi = \prt{\omega, \alpha, I_a, D_{5-95}, t_{mid}, \omega_{mid}, \omega', \zeta}$.

\begin{table}[!ht]
\caption{Marginal distributions of the Bouc Wen model parameters}
\centering
\begin{tabular}{ccccc}
\hline
Parameter & Distribution & Support & Mean & Standard deviation  \\
\hline
$\omega$ (rad/s) & Uniform & $[5.373,\, 6.567]$  & 5.97 & 0.3447\\
$\alpha$ (1/m) & Uniform & $[45,\, 55]$  & 50 & 2.887\\
\hline
\end{tabular}
\label{tab5.3.1}
\end{table}

We first build the metamodel for representing the velocity time histories $v(t)$ of the oscillator. 
200 simulations are conducted with 200 samples of the input parameters generated by Latin hypercube sampling. The system of ODEs are solved by means of the Matlab solver \textit{ode45} (explicit Runge-Kutta method with relative error tolerance $1 \, \times 10^{-3}$) with the total duration $T=30$~s and time step $\di t=0.005$~s as in the previous example. In the first place, a NARX model structure is chosen, in which the model terms are $g_i(t) = x(t-i)^l \abs{v(t-1)}^m$ and $g_i(t) =  v(t-j)^l \abs{v(t-1)}^m$ with $l= 0,1$, $m = 0,1$, $j = 1 \enum 4$, $i = 0 \enum 4$. 
The use of absolute terms has proven effective in capturing the hysteretic behaviour of nonlinear systems in \citep{Spiridonakos2015}.
The initial NARX model contains 19 terms in total.

\begin{figure}[ht!]
	\centering
	\subfigure[$\vexi = (6.24, 53.64, 0.04,    6.74,    7.54,   17.22,    0.08,    0.26)$]
	{
	\includegraphics[width=0.45\linewidth,height = 5.7cm]{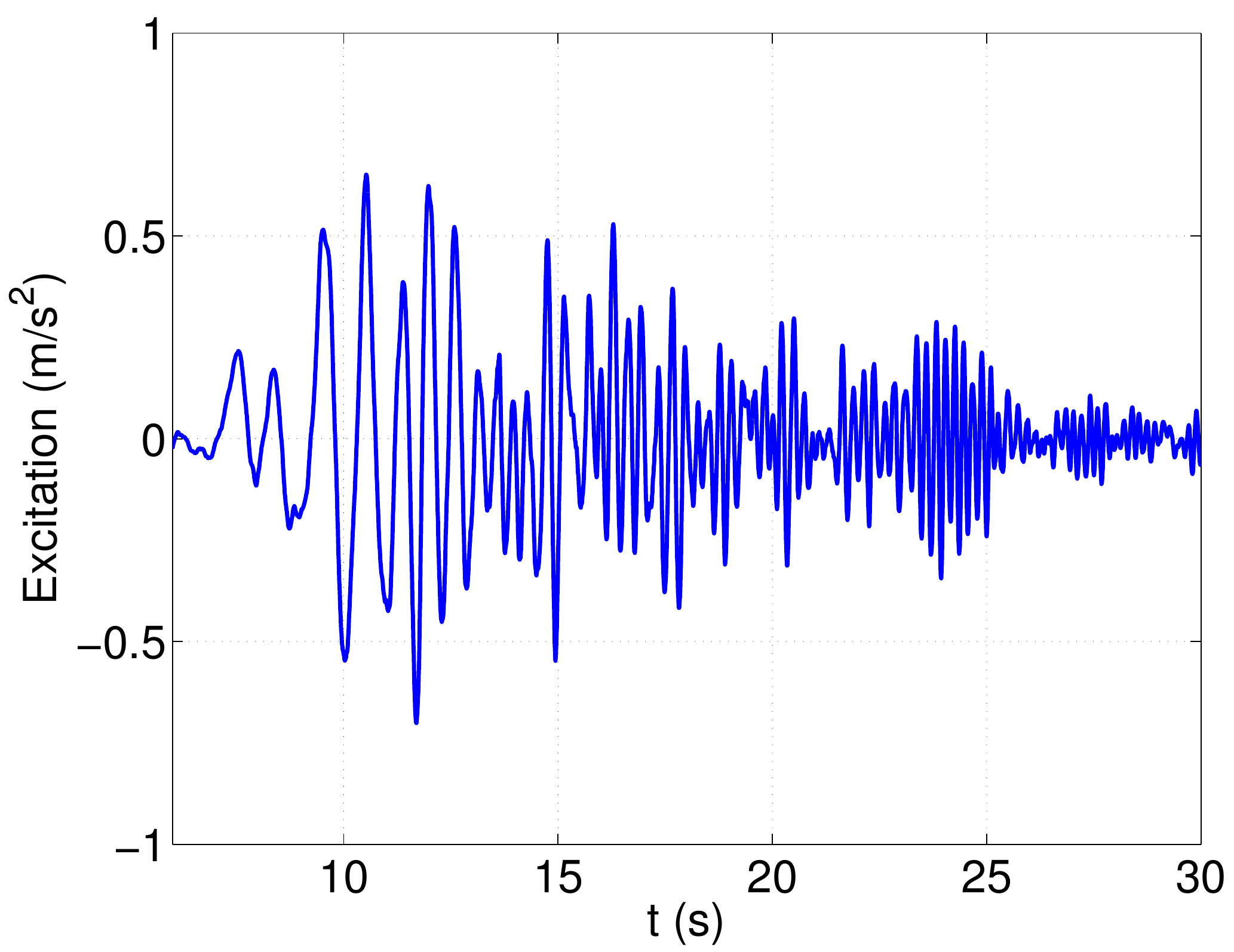}
	}
	\subfigure[Nonlinear behavior]
	{
	\includegraphics[width=0.45\linewidth]{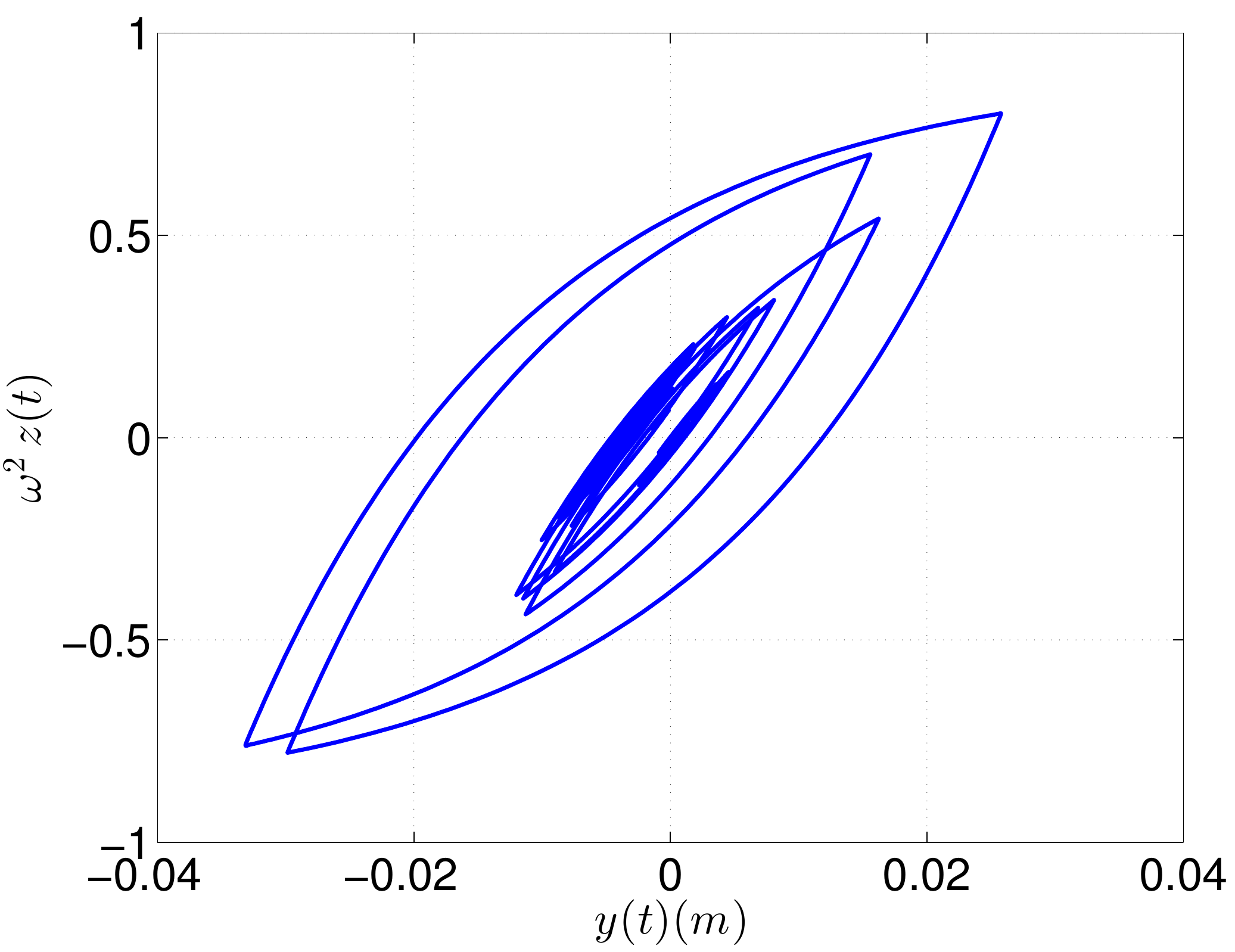}
	}
	\subfigure[Displacement]
		{
		\includegraphics[width=0.45\linewidth]{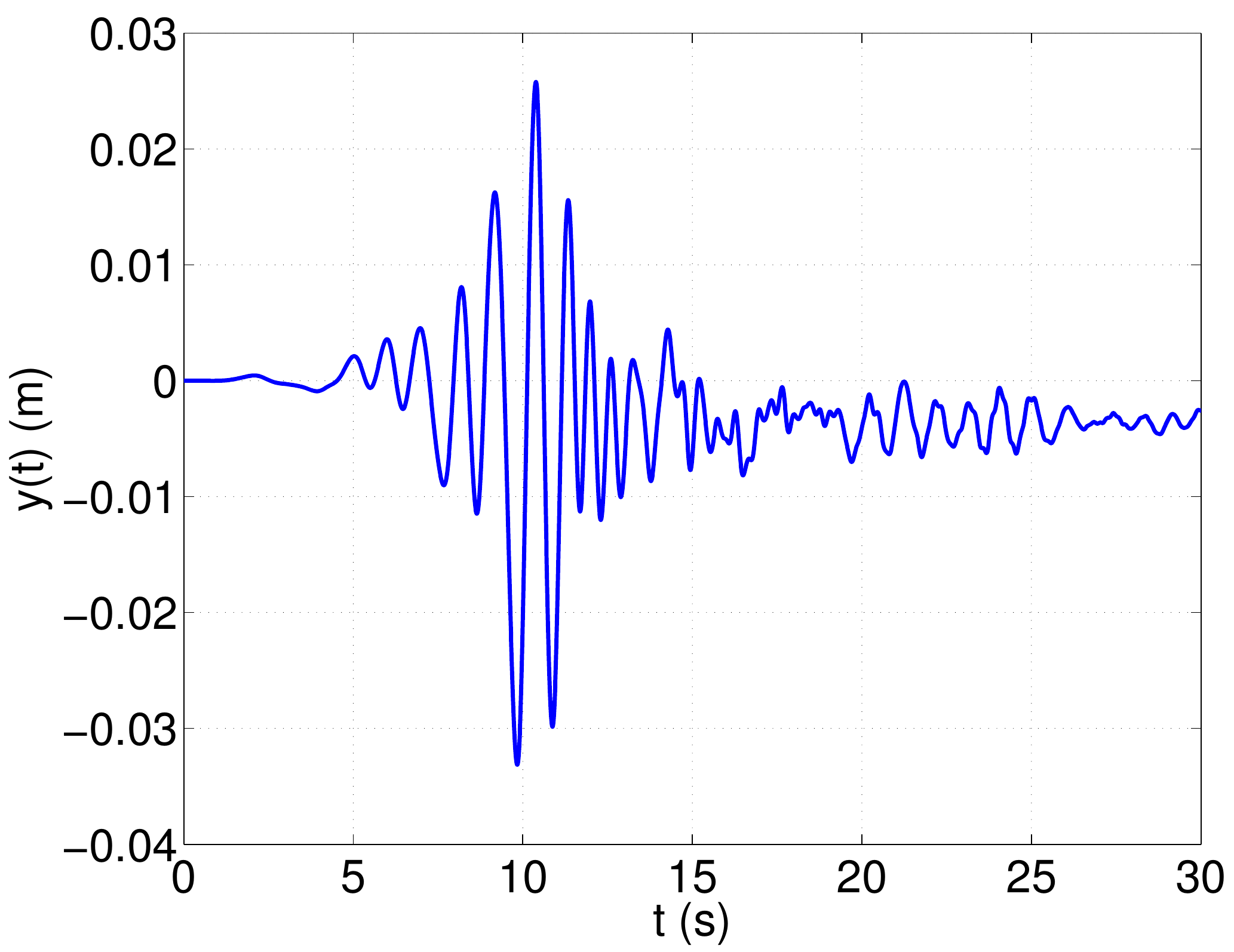}
		}
	\subfigure[Velocity]
			{
			\includegraphics[width=0.45\linewidth]{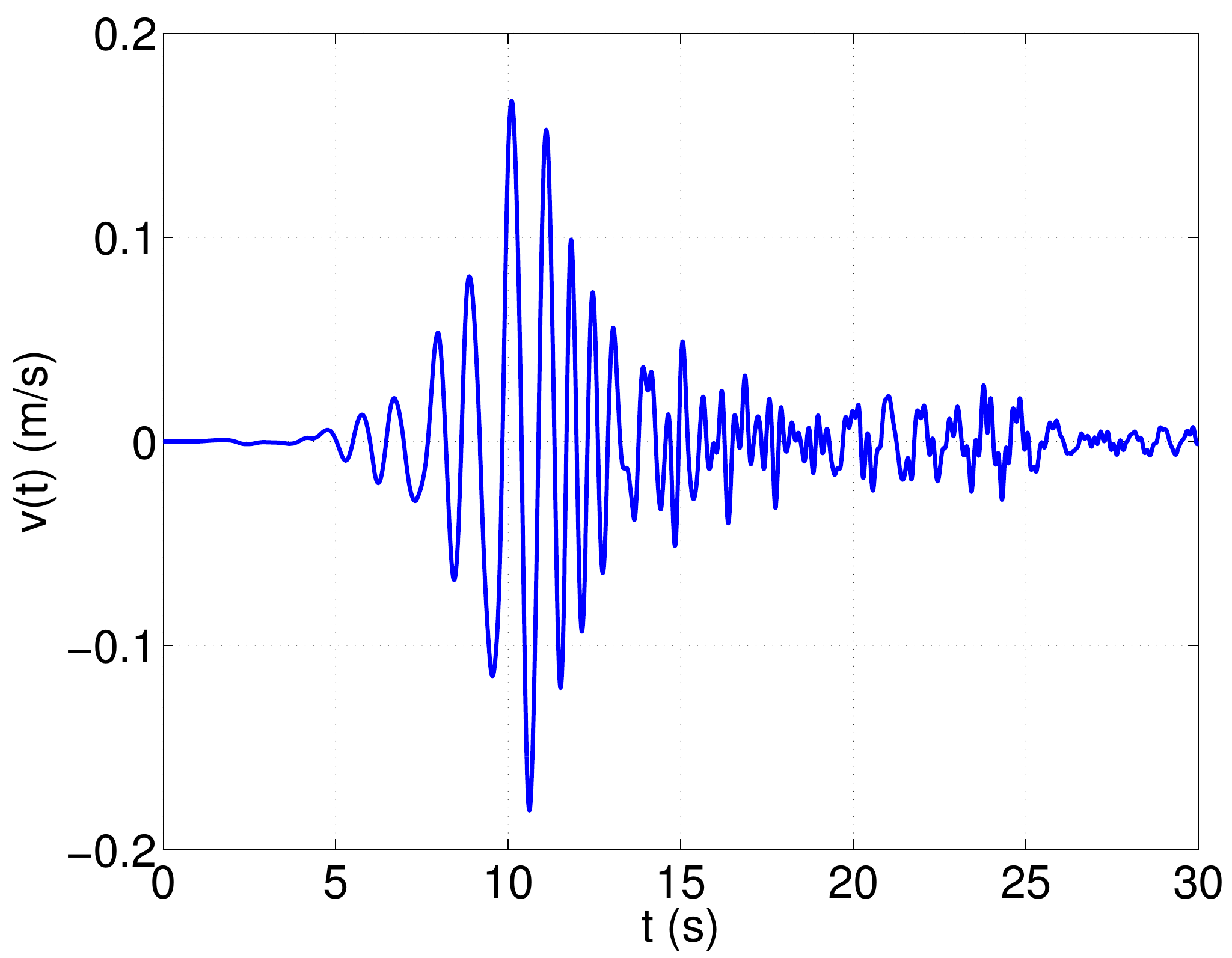}
			}
	\caption{Bouc Wen oscillator -- Experiment leading to the selected NARX structure.}
	\label{fig5.3.2}
\end{figure}

Next, the candidate NARX models were computed. For this purpose, we selected the simulations with maximum velocity exceeding a large threshold, \ie $\max(\abs{v(t)}) > 0.25$~m/s and obtained 15 experiments. LARS was applied to the initial full NARX model to detect the most relevant NARX terms constituting a candidate NARX model from each simulation previously selected. This procedure resulted in 11 candidates in total.
OLS (Eq.~\eqref{eqOLSsolution}) is used to determine the NARX coefficients corresponding to each NARX candidate model for all the simulations. To evaluate the accuracy of the NARX candidate, Eq.~\eqref{eqSE} is used to compute the error indicators.
The most appropriate NARX model achieves a mean relative error $\bar{\epsilon}= 6.27 \times \,10^{-4}$ over 200 experiments and contains 12 terms, namely constant term, $x(t-4)$, $x(t-4) \, \abs{v(t-1)}$, $x(t-3)$, $x(t-3) \, \abs{v(t-1)}$, $x(t-2)$, $x(t-1)$, $x(t)$,  $v(t-4)$, $v(t-4)\, \abs{v(t-1)}$, $v(t-3) \, \abs{v(t-1)}$, $v(t-1)$. \figref{fig5.3.2} depicts the experiment from which the most appropriate NARX model is selected.
Note that the nonlinear behaviour is noticeable and the oscillator exhibits a residual displacement after entering the domain of nonlinearity.

Then we represented the NARX coefficients by sparse PCEs. The optimal polynomial of order $p=2$ was found adaptively with maximum interaction order $r=2$ and truncation parameter $q = 1$ so that the resulting PC-NARX model led to the smallest error when reconstructing the responses in the ED.
The PCEs of the NARX coefficients have LOO errors smaller than $1.68\, \times 10^{-4}$. 
The PC-NARX model of the velocity was obtained and used for predicting the velocity on the validation set.
The displacement time history is then obtained by integration.

\begin{figure}[ht!]
	\centering
	\subfigure[First example trajectory $\vexi = (6.35, 46.54, 0.05,    6.53,    5.47,   55.16,   -1.02,    0.31)$]
		{
		\includegraphics[width=0.45\linewidth]{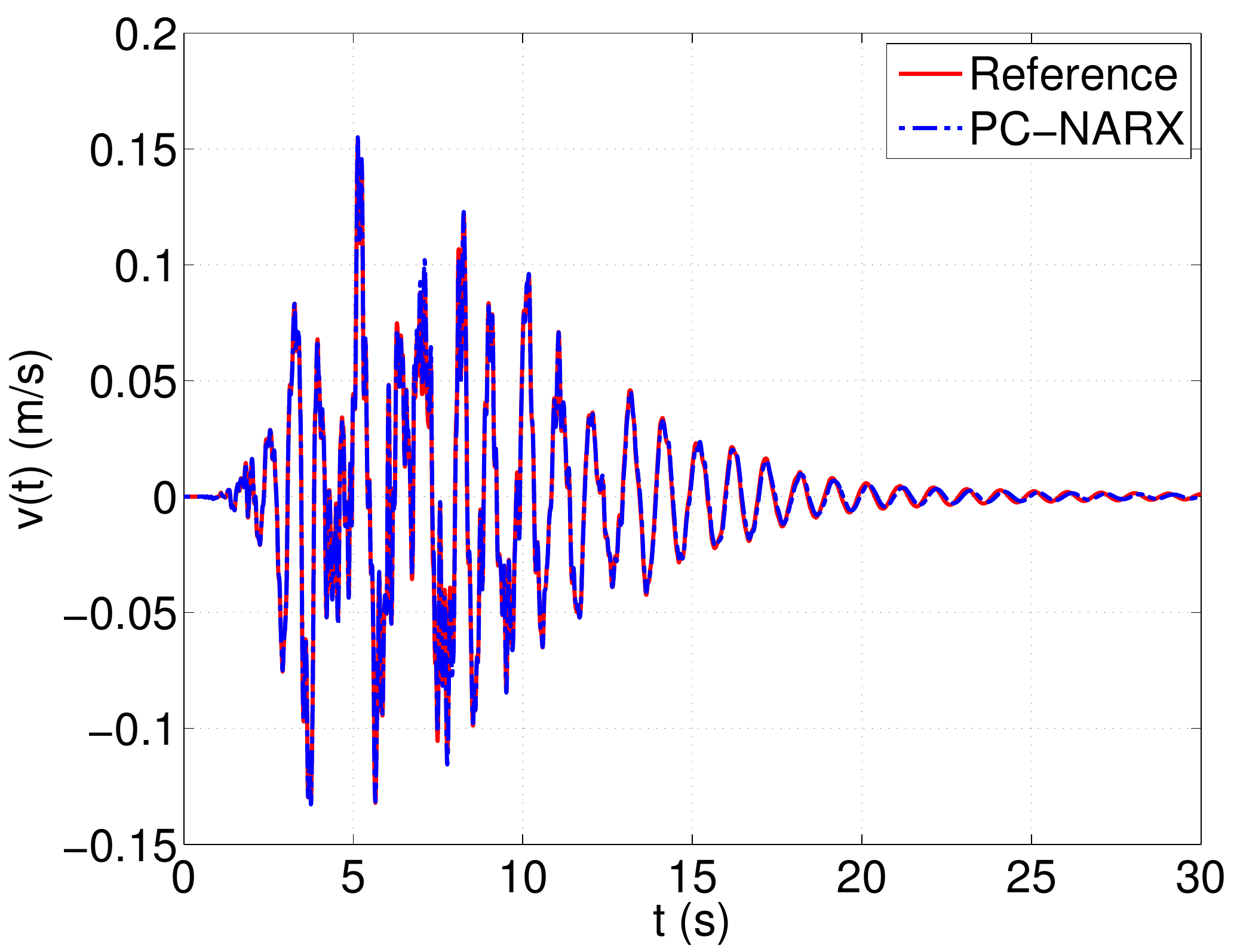}
		}
	\subfigure
		{
		\includegraphics[width=0.45\linewidth]{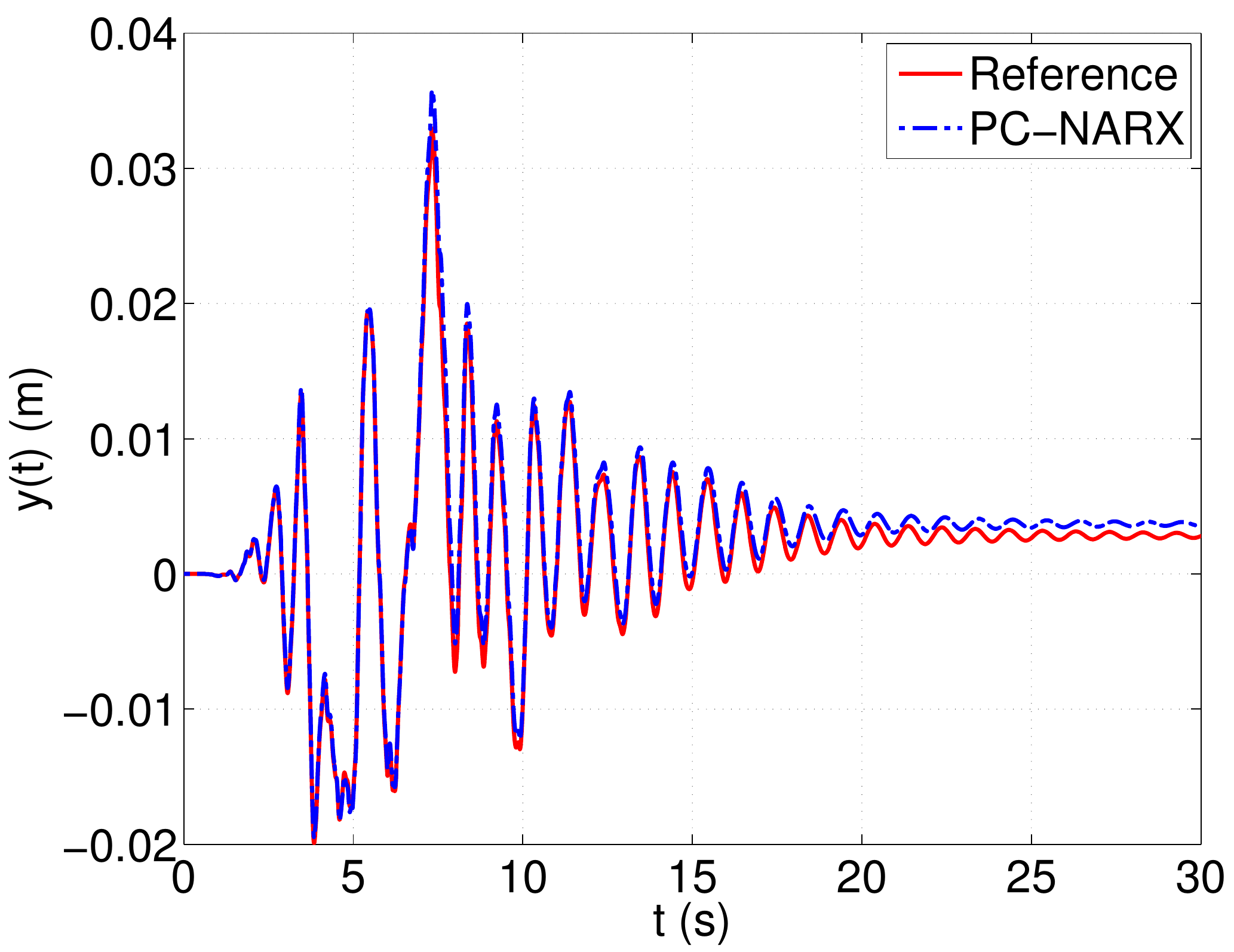}
		}
		
	\subfigure[Second example trajectory $\vexi = (6.32, 48.24, 0.05,    9.08,    4.53,   19.76,   -0.11,    0.28)$]
		{
		\addtocounter{subfigure}{-1}
		\includegraphics[width=0.45\linewidth]{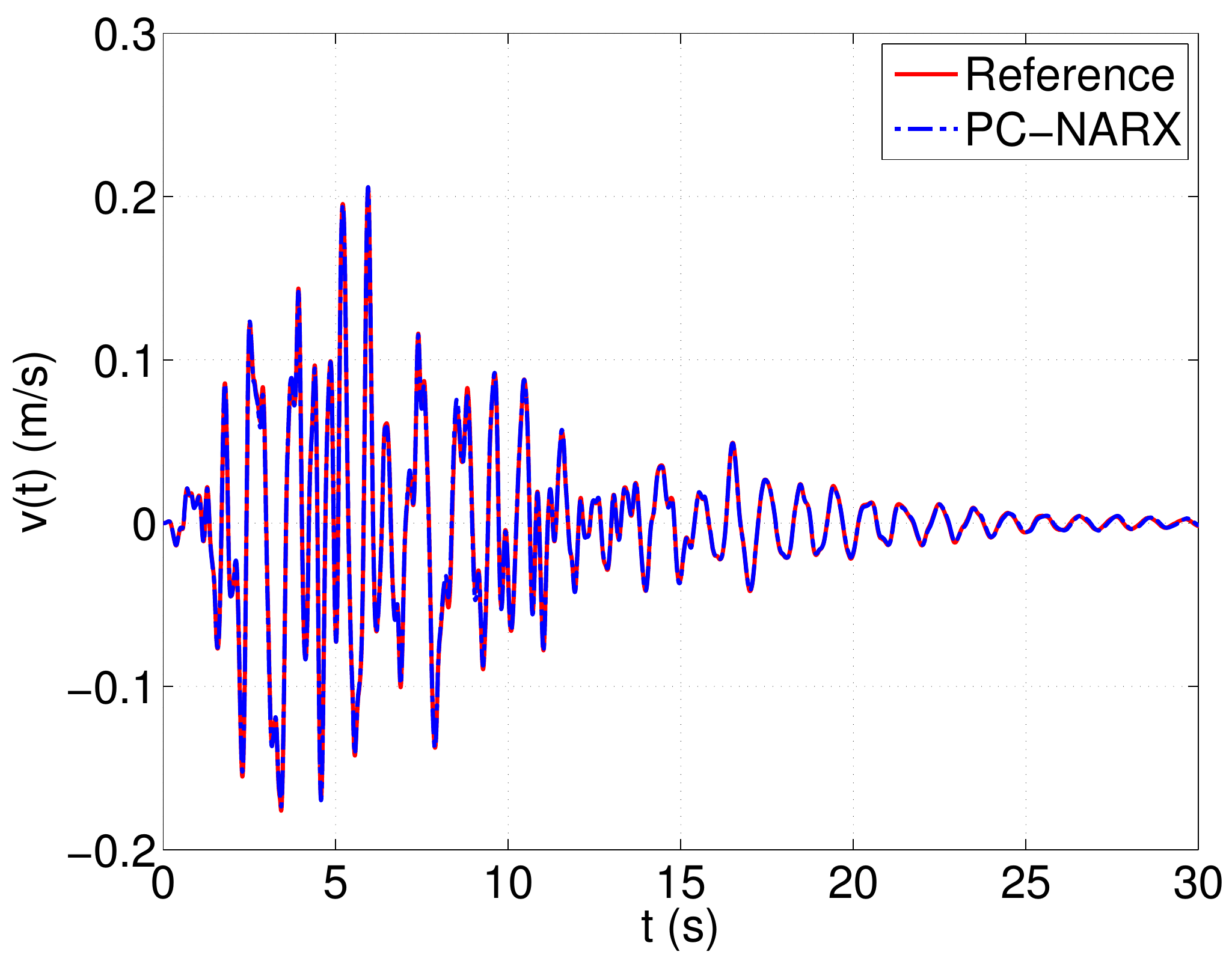}
		}
	\subfigure
		{
		\includegraphics[width=0.45\linewidth]{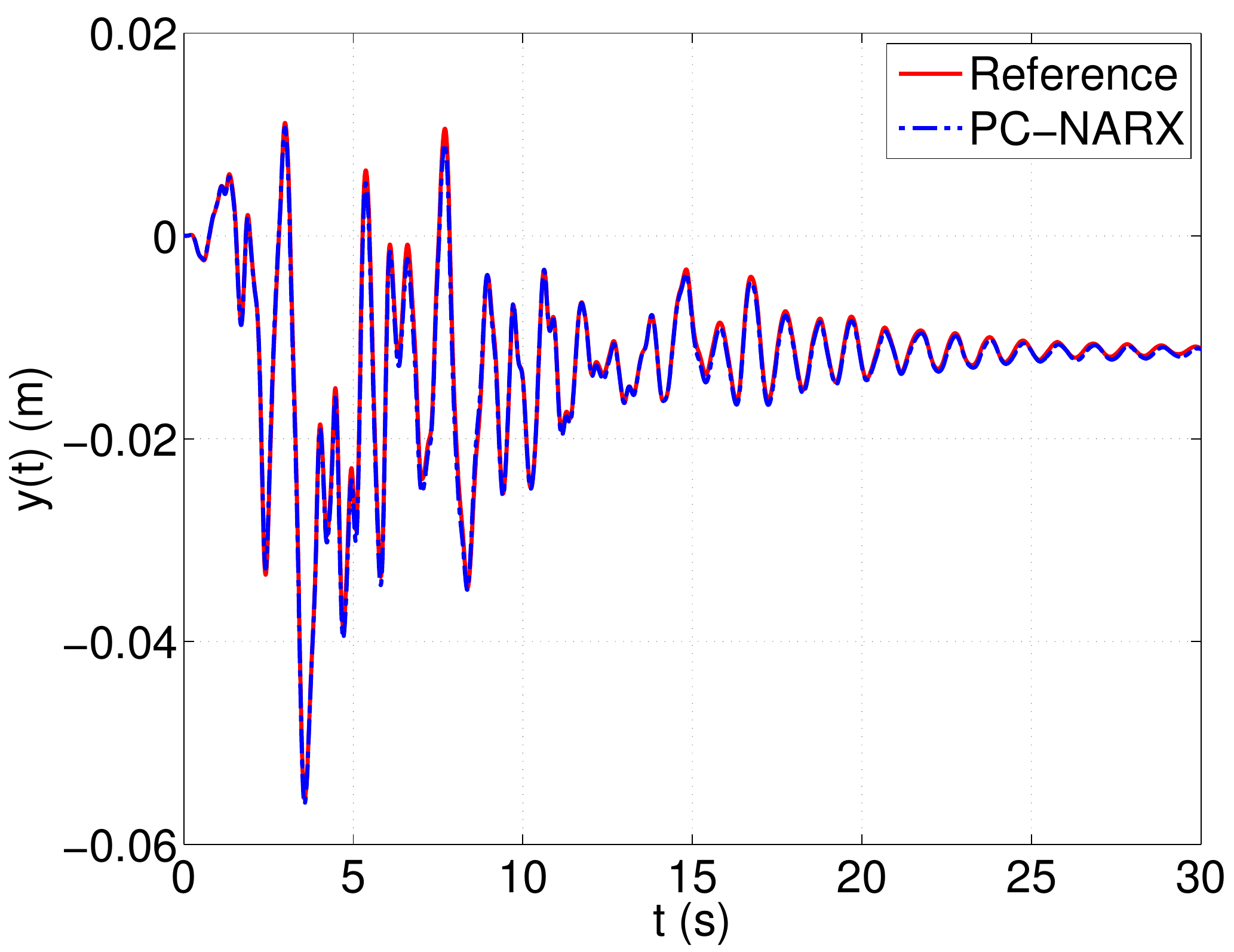}
		}
		\caption{Bouc Wen oscillator -- Two particular trajectories of velocity $v(t)$ and displacement $y(t)$ and their predictions by means of PC-NARX.}
	\label{fig5.3.4}
\end{figure}

\figref{fig5.3.4} depicts two specific velocity and displacement trajectories due to distinct validation sets of parameters.
One observes that the velocity trajectories are relatively well predicted by PC-NARX. Indeed, the mean relative error over $10^4$ validations is $\bar{\epsilon}_{val}= 1.82\times 10^{-2}$ and only $3\%$ of those time-histories have a relative error $\epsilon_{val, i}$ exceeding $0.1$.
Despite the high accuracy of the PC-NARX model for the velocity, the predicted displacements exhibit some slight discrepancies with respect to the actual trajectories. The inaccuracies are slightly large in the later phase of the considered time domain due to the occurrence of residual displacements, while remaining acceptable.

\begin{figure}[ht!]
	\centering
	\subfigure[Standard deviation of velocity $v(t)$]
	{
	\includegraphics[width=0.45\linewidth]{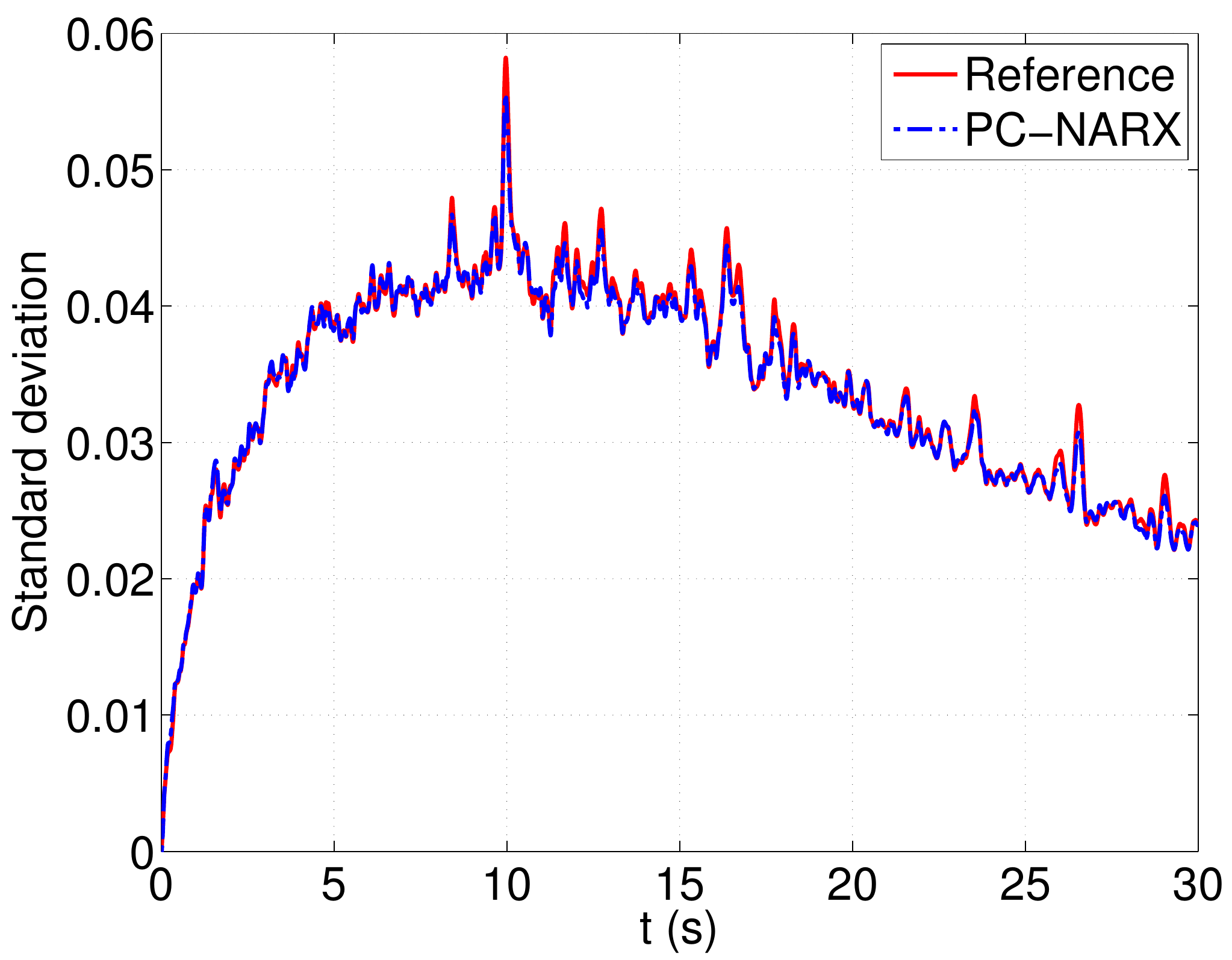}
	}
		\subfigure[Standard deviation of displacement $y(t)$]
		{
		\includegraphics[width=0.45\linewidth]{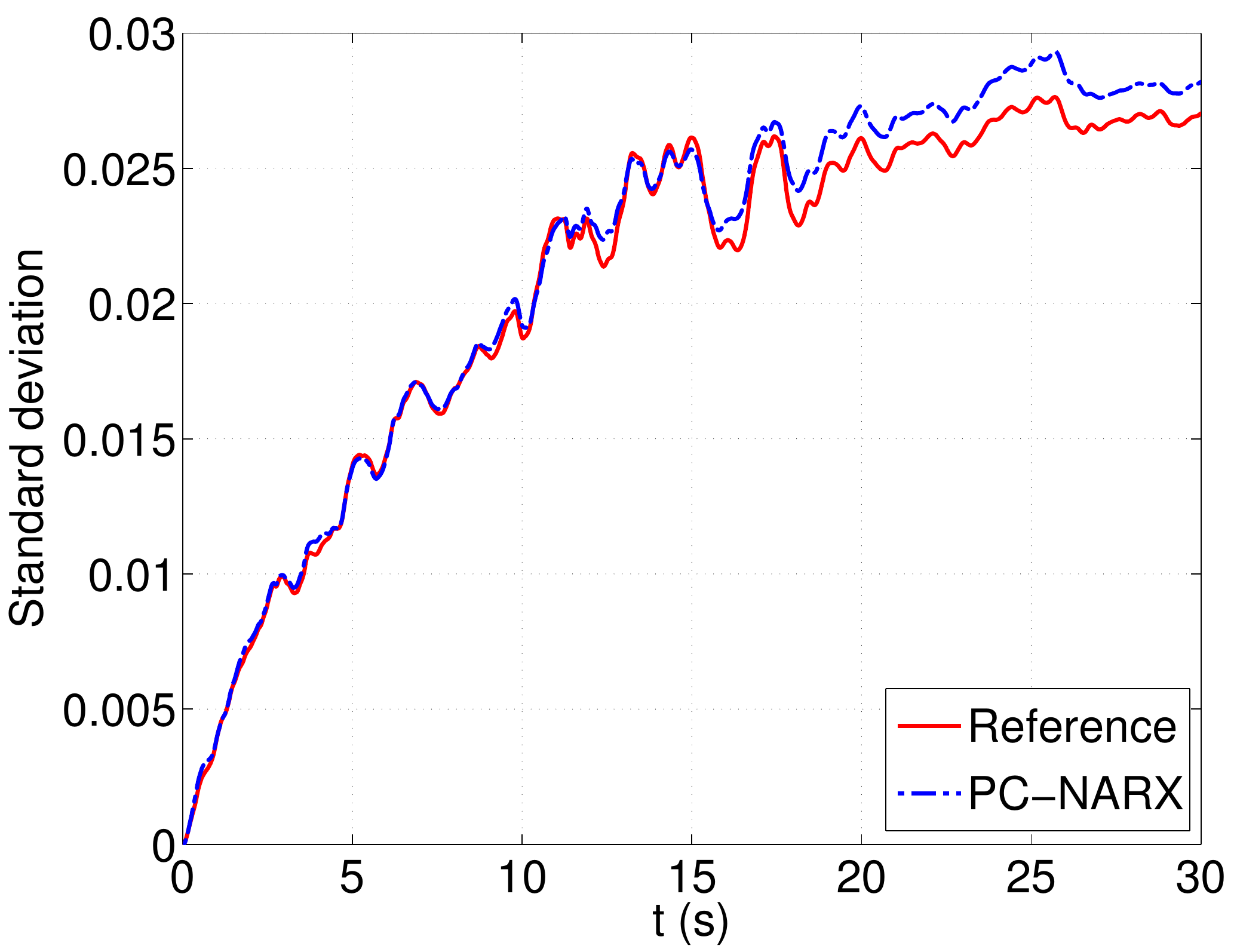}
		}
		
	\caption{Bouc Wen oscillator -- Standard deviation trajectories of the responses. }
	\label{fig5.3.5}
\end{figure}

\figref{fig5.3.5} compares the time-dependent standard deviation of the two response quantities predicted by PC-NARX with those obtained from Monte Carlo simulation. The high frequency content is observed in the evolutionary response statistics, which was also reported in the literature when applying Monte Carlo simulation to a nonlinear structure \citep{Kougioumtzoglou2013}.The standard deviation of the velocity is remarkably well captured by PC-NARX (relative error $\epsilon_{val, std} = 0.39 \times 10^{-2}$). For the displacement, the standard deviation tends to increase in time, which is different from the Duffing oscillator that does not exhibit residual displacement. The discrepancy between the prediction and the actual time histories is also increasing in time. However, the resulting relative error remains rather small ($\epsilon_{val, mean} = 1.57 \times 10^{-2}$).

\begin{figure}[ht!]
	\centering
	\subfigure[Maximal values of velocity]
	{
	\includegraphics[width=0.45\linewidth]{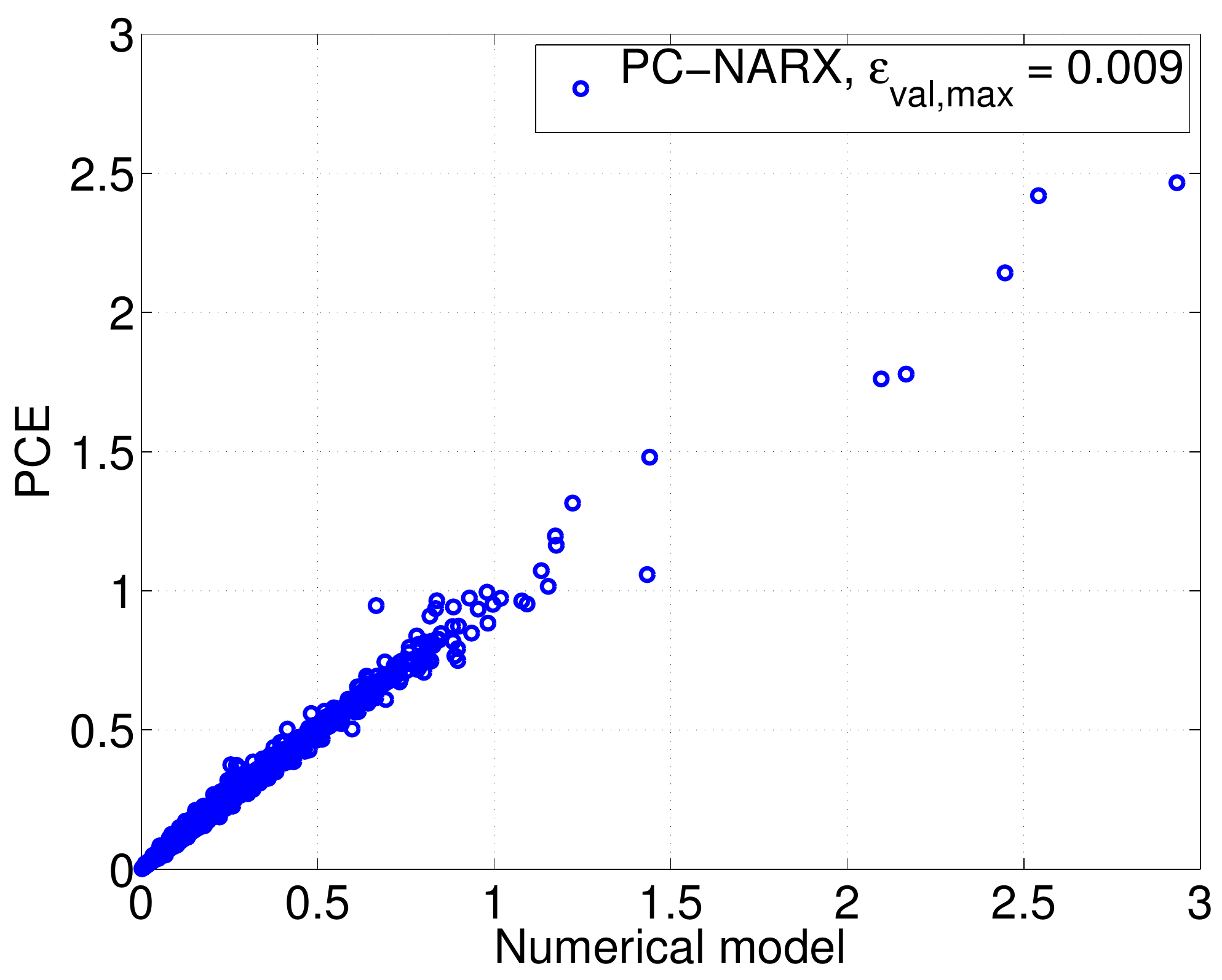}
	}
	\subfigure[Probability density function of maximal values of velocity]
	{
	\includegraphics[width=0.45\linewidth, height =5.95cm]{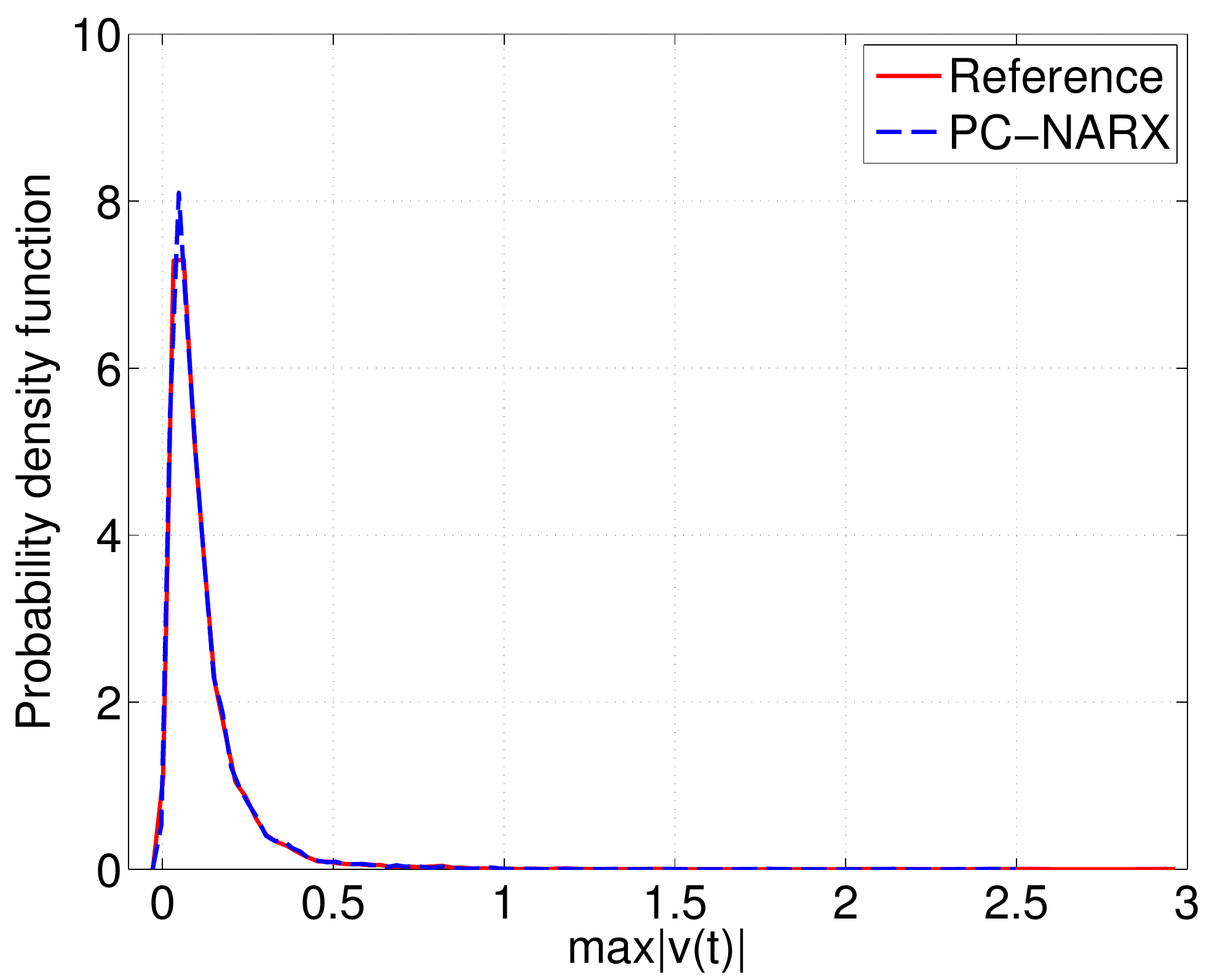}
	\label{fig5.3.6.2}
	}
	
	\subfigure[Maximal values of displacement]
		{
		\includegraphics[width=0.45\linewidth]{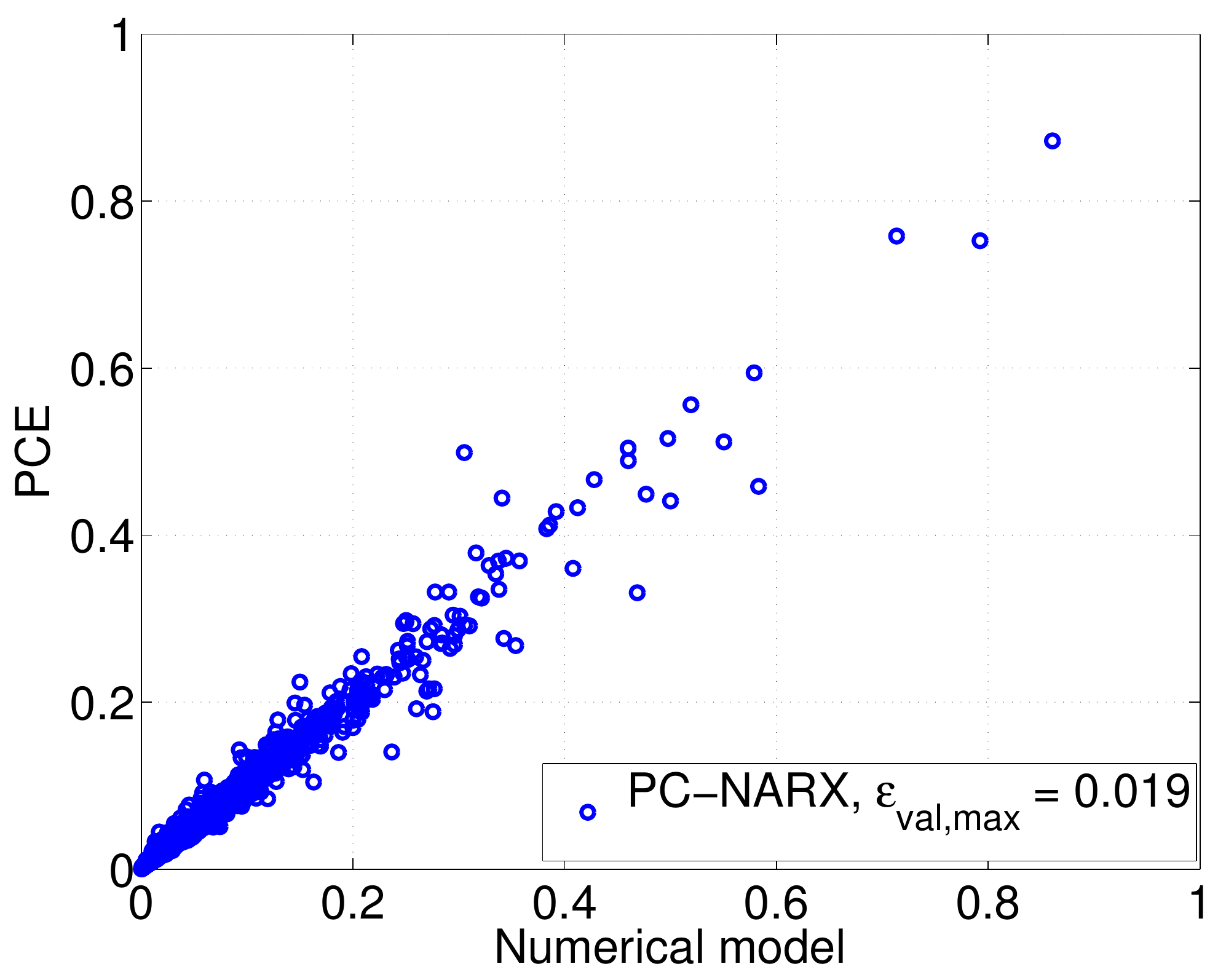}
		}
		\subfigure[Probability density function of maximal values of displacement]
		{
		\includegraphics[width=0.45\linewidth, height =5.95cm]{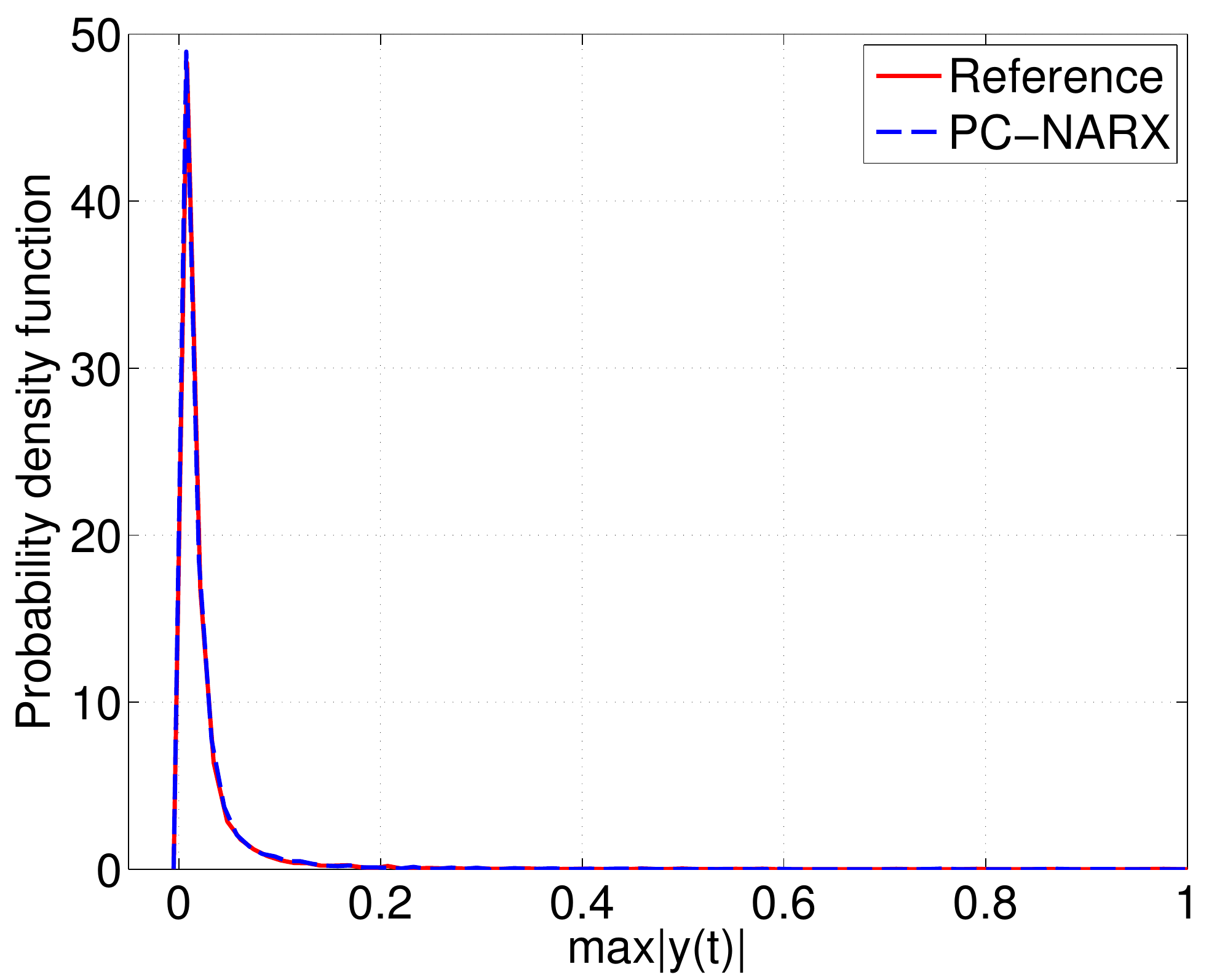}
		\label{fig5.3.6.4}
		}
		
	\caption{Bouc Wen oscillator -- Maximal responses computed by numerical solver and predicted by PC-NARX model.}
	\label{fig5.3.6}
\end{figure}

\figref{fig5.3.6} compares the maximum values of velocity and displacement predicted by PC-NARX with those values computed by the numerical solver. Despite the complexity of the problem, the predictions are remarkably consistent with the true values, with sufficiently small validation errors of $\epsilon_{val, \max(|v(t)|)} = 9 \times 10^{-3}$ and $\epsilon_{val, \max(|y(t)|)} = 1.9 \times 10^{-2}$. The accurate predictions allow one to obtain the probability density functions of the maximum responses that are in good agreement with the reference functions, as shown in \figref{fig5.3.6.2} and \figref{fig5.3.6.4}.

\section{Discussion}

In this section, we discuss certain features of the PC-NARX approach.
It is clear that PC-NARX outperforms traditional time-frozen PCEs.
The latter are reported to fail in representing the random response of dynamical systems.
The reason is that PCEs are designed for propagating uncertainties, but not for capturing the evolutionary behaviour in time domain. In order to use PCEs in this context, one should utilize a specific tool
to tackle the dynamical aspect of the problem. For instance, one can use the available system of equations describing the model in an intrusive way. This approach is, however, also known to deteriorate in long term integration due to the accumulation of error in time \citep{Wan2006,Ghosh2007}.
One also apprehends the remarkable performance of PC-NARX approach, which handles separately the two aspects of the problem, namely dynamics and uncertainties with the specific tools of NARX modelling and PCEs, respectively. 

PC-NARX can also be considered to belong to a class of approaches where a pre-processing of the response time histories is conducted before applying PCEs. The response trajectories are first projected onto a different space, for instance phase-space \citep{Desai2013}, transformed time space \citep{LeMaitre2009} or space of NARX functions \citep{Spiridonakos2015}. In the case of PC-NARX, the NARX functions form a ``coordinate system'' that moves along the trajectory. Due to the projection of the trajectories on the \emph{suitably selected} set of basis functions, the projection coefficients become smooth functions of the uncertain parameters, therefore they can be represented effectively by low order PCEs.

The PC-NARX approach also has its limitations. It uses data observed in discrete time. When the recorded data is sparse in the time domain, \ie the time step is too large, the mechanism that relates the current state with the previous state and excitation cannot be revealed.
In addition, the recursive form of the PC-NARX formulation renders it more difficult for post-processing. For instance, closed-form solutions for time-dependent statistics and sensitivity analysis are currently not available.

In general, we are convinced that approaches which make use of interdisciplinary tools such as PC-NARX are promising. In other words, PCEs should be combined with tools designed for predicting non-linear time series in uncertainty quantification 
of stochastic dynamical systems.






\section{Conclusions}

Polynomial chaos expansions are a versatile tool for uncertainty quantification, which have proven
effective in various fields. However, the propagation of uncertainties in stochastic dynamical systems
remains a challenging issue for PCEs as well as all other metamodelling techniques.
It is widely known that PCEs fail to capture the long-term dynamics of the underlying system. Therefore, a specialized tool must be used to handle this aspect of the problem.
Nonlinear autoregressive with exogenous input (NARX) models are universally used in the field of system identification for revealing
the dynamical behaviour from observed time series. 
The combination of NARX and PCEs have been proposed recently and shown great effectiveness in the context of stochastic dynamics.

In this paper we introduced the least angle regression (LARS) technique for building PC-NARX models of nonlinear
systems with uncertainties subject to stochastic excitations. 
In particular, the approach consists in solving two linear regression problems.
LARS proves suitable for selecting both the appropriate NARX and PCE models, which plays a crucial role in the approach. The LARS-based PC-NARX approach is applied to predict the response time histories of three benchmark engineering models subject to non-stationary stochastic excitation with remarkable accuracy. 
It is important to observe that, to our knowledge, no generic non-intrusive tools based on PCEs were available so far to build surrogate models of dynamical system responses based on a set of computer experiment.

Investigation on the application of the proposed approach to more complex problems is ongoing, with
particular interests being focused on the seismic analysis of structures.

\nocite{SudretHDR}

\bibliographystyle{chicago}
\bibliography{biblioRSUQ}
\end{document}